\documentclass[11pt,a4paper]{article}
\usepackage{jcappub}
\pdfoutput=1

\title{Fitting Methods for Baryon Acoustic Oscillations in the Lyman-$\alpha$ Forest Fluctuations in BOSS Data Release 9}

\author[a]{David Kirkby,}
\author[a]{Daniel Margala,}
\author[b]{An\v{z}e Slosar,}

\author[c]{Stephen~Bailey,}
\author[d]{Nicol\'as G. Busca,} 
\author[e]{Timoth\'ee Delubac,}
\author[e]{James Rich,}

\author[a]{Michael~Blomqvist,}
\author[f]{Joel R. Brownstein,}
\author[c]{Bill Carithers,}
\author[g]{Rupert A.C. Croft,}
\author[f]{Kyle S. Dawson,}
\author[h,c]{Andreu Font-Ribera,}
\author[i,j]{Jordi Miralda-Escud\'{e},}
\author[k]{Adam D. Myers,}
\author[l]{Robert C. Nichol,}
\author[e]{Nathalie Palanque-Delabrouille,}
\author[m,n]{Isabelle P\^aris,}
\author[m]{Patrick Petitjean,}
\author[e]{Graziano Rossi}
\author[c]{David J. Schlegel,}
\author[o,p]{Donald P. Schneider,}
\author[q,r]{Matteo Viel,}
\author[s]{David H. Weinberg,}
\author[e]{Christophe Y\`eche}

\affiliation[a]{Department of Physics and Astronomy, University of California, Irvine, 92697, USA}
\affiliation[b]{Brookhaven National Laboratory, Blgd 510, Upton NY 11375, USA}
\affiliation[c]{Lawrence Berkeley National Laboratory, 1 Cyclotron  Road, Berkeley, CA 94720, USA}
\affiliation[d]{APC, Universit\'{e} Paris Diderot-Paris 7, CNRS/IN2P3, CEA,  Observatoire de Paris, 10, rueA. Domon \& L. Duquet,  Paris, France}
\affiliation[e]{CEA, Centre de Saclay, IRFU,  F-91191 Gif-sur-Yvette, France}
\affiliation[f]{Department of Physics and Astronomy, University of Utah, 115 S 1400 E, Salt Lake City, UT 84112, USA}
\affiliation[g]{Bruce and Astrid McWilliams Center for Cosmology, Carnegie Mellon University, Pittsburgh, PA 15213, USA}
\affiliation[h]{Institute of Theoretical Physics, University of Zurich, 8057 Zurich, Switzerland }
\affiliation[i]{Instituci\'{o} Catalana de Recerca i Estudis  Avan\c{c}ats, Barcelona, Catalonia}
\affiliation[j]{Institut de Ci\`{e}ncies del Cosmos, Universitat de Barcelona/IEEC, Barcelona 08028, Catalonia}
\affiliation[k]{Department of Physics and Astronomy, University of Wyoming, Laramie, WY 82071, USA}
\affiliation[l]{Institute of Cosmology and Gravitation, Dennis Sciama Building, University of Portsmouth, Portsmouth, PO1 3FX, UK}
\affiliation[m]{Universit\'e Paris 6 et CNRS, Institut d'Astrophysique de Paris, 98bis blvd. Arago, 75014 Paris, France}
\affiliation[n]{Departamento de Astronom\'ia, Universidad de Chile, Casilla 36-D, Santiago, Chile}
\affiliation[o]{Department of Astronomy and Astrophysics, The Pennsylvania State University, University Park, PA 16802, USA}
\affiliation[p]{Institute for Gravitation and the Cosmos, The Pennsylvania State University, University Park, PA 16802, USA}
\affiliation[q]{INAF, Osservatorio Astronomico di Trieste, Via G. B. Tiepolo 11, 34131 Trieste, Italy}
\affiliation[r]{INFN/National Institute for Nuclear Physics, Via Valerio 2, I-34127 Trieste, Italy.}
\affiliation[s]{Department of Astronomy, Ohio State University, 140 West 18th Avenue, Columbus, OH 43210, USA}

\emailAdd{dkirkby@uci.edu}

\abstract{We describe fitting methods developed to analyze fluctuations in the Lyman-$\alpha$ forest and measure the parameters of baryon acoustic oscillations (BAO). We apply our methods to BOSS Data Release 9. Our method is based on models of the three-dimensional correlation function in physical coordinate space, and includes the effects of redshift-space distortions, anisotropic non-linear broadening, and broadband distortions. We allow for independent scale factors along and perpendicular to the line of sight to minimize the dependence on our assumed fiducial cosmology and to obtain separate measurements of the BAO angular and relative velocity scales. Our fitting software and the input files needed to reproduce our main BOSS Data Release 9 results are publicly available.}


\begin{document}
\maketitle

\section{Introduction}

The surprising discovery~\cite{RIFIET98,PEADET99} of accelerating expansion in the current universe reveals that either some form of dark energy is driving the expansion or else that our theory of gravity is incomplete on the largest scales (see reference~\cite{2012arXiv1201.2434W} for a recent review). The length scale of baryon acoustic oscillations (BAO) imprinted at the moment of baryon-photon decoupling provides a standard ruler that has been well measured in the temperature anisotropies of the cosmic microwave background~\cite{2011ApJS..192...18K,2012arXiv1212.5226H} (CMB) and in the number-density fluctuations of galaxies at $z < 1$~\cite{2001MNRAS.327.1297P, 2001ApJ...555...68M, 2005ApJ...633..560E, 2005MNRAS.362..505C, 2006A&A...459..375H, 2007MNRAS.378..852P, 2007MNRAS.374.1527B, 2007MNRAS.381.1053P, 2008ApJ...676..889O, 2009MNRAS.399.1663G, 2010ApJ...710.1444K, 2010MNRAS.401.2148P, 2010MNRAS.404...60R, 2011MNRAS.418.1707B, 2011MNRAS.416.3017B, 2012ApJ...761...13S, 2012arXiv1202.0090P, 2012arXiv1203.6594A, 2012arXiv1211.2616B}. The Baryon Oscillation Spectroscopic Survey~\cite{2013AJ....145...10D} (BOSS) of the third generation of the Sloan Digital Sky Survey~\cite{2011AJ....142...72E} (SDSS-III) recently announced its ninth data release~\cite{2012ApJS..203...21A} (DR9), including an unprecedented number of high-redshift quasar spectra. The BAO feature is imprinted in the correlated fluctuations of intergalactic Lyman-$\alpha$ absorption of the light from these quasars, enabling us to measure the BAO standard ruler at redshifts $z \simeq 2.4$, during the predicted era of matter-dominated deceleration.

An optimal extraction of the BAO signal from a large sample of Lyman-$\alpha$ forest pixels requires new fitting techniques beyond those previously developed to fit the large-scale clustering of galaxies~\cite{2012arXiv1202.0091X}. For example, the non-uniform sampling of the absorption field, large expected redshift-space distortions, and relatively large depth suggest that estimates of the correlation function or power spectrum should be fit to three-dimensional models (rather than low-order multipoles at fixed redshift) expressed directly in terms of the physical absorption wavelengths and angular separations between lines of sight. We describe here new fitting techniques developed specifically for a near-optimal analysis of the DR9 correlation-function estimates described in a companion paper~\cite{Slosar2013}, and highlight the challenges and lessons learned. We are also making our fit input files and fitting code publicly available as a companion to this paper, so that readers may reproduce the main results presented here and in ref.~\cite{Slosar2013}.

The outline of our paper is as follows. In Section~\ref{sec:models} we define our constituent models for linear theory with redshift-space distortions, non-linear effects, and redshift evolution. We also describe our parametrizations of possible deviations between our assumed fiducial cosmology\footnote{We assume a flat $\Lambda$CDM universe with $\Omega_\Lambda = 0.73$, $h = 0.7$, $\Omega_{\text{b}} h^2 = 0.0227$, and $n_s = 0.97$ throughout.} and the cosmology preferred by our data, and of broadband distortion of our correlation-function estimate introduced by analysis systematics. Finally, we introduce two data-reduction techniques using interpolated models. In Section~\ref{sec:fitting}, we describe the DR9 fitting inputs which consist of $N = 1512$ correlation-function estimates on a three-dimensional physical coordinate grid, accompanied by an initial estimate of their covariance in a block diagonal form that reduces the number of non-zero elements from $N(N+1)/2 \simeq 1,114$K to $\simeq 64$K. Next, we describe a novel method for internally validating and refining our initial covariance estimate, necessitated by a lack of available simulated mock statistics. In Section~\ref{sec:results}, we present our results on the fitting method and what it teaches us about the BOSS DR9 Lyman-$\alpha$ forest dataset. We discuss the expected parameter sensitivities and relative contributions from different regions of the three-dimensional separation space, and present model-independent data reduction results. Cosmological fitting results for DR9 are presented in the companion paper ref.~\cite{Slosar2013}. We conclude in Section~\ref{sec:discussion} with a discussion of the main lessons learned and our plans for future development. Appendix~\ref{sec:public} provides details on public access to our fit inputs and fitting code.

\section{Models and Parameters}
\label{sec:models}

We model a measurement of the correlation function
\begin{equation}
\xi(r,\mu,z) \equiv \langle \delta(\mathbf{s}_1) \delta(\mathbf{s}_2) \rangle - \langle \delta(\mathbf{s}_1) \rangle \langle \delta(\mathbf{s}_2) \rangle
\end{equation}
where the ensemble averages are taken over realizations of a (possibly biased) tracer $\delta(\mathbf{s})$ of the large-scale distribution of matter in redshift space $\mathbf{s}$, with $(r,\mu,z)$ defined via\footnote{We use the notation $\mu \equiv \hat{\mathbf{z}}\cdot \hat{\mathbf{r}}$ and $\mu_k \equiv \hat{\mathbf{z}}\cdot \hat{\mathbf{k}}$.}
\begin{equation}
|\mathbf{s}_2 - \mathbf{s}_1| = r \quad , \quad
|\mathbf{s}_2| - |\mathbf{s}_1| = \mu\cdot r \quad , \quad
\frac{1}{2}\,|\mathbf{s}_1 + \mathbf{s}_2| = c \int_0^z \, \frac{dz'}{H_{\text{fid}}(z')}
\end{equation}
for some fiducial cosmology with Hubble function $H_{\text{fid}}(z)$.  The model combines a cosmological prediction $\xi_{\text{cosmo}}$ with a parametrization of possible multiplicative and additive broadband distortions introduced by the analysis method.

\subsection{Physical Coordinates}
\label{sec:physical-coords}

\begin{figure}[htb]
\begin{center}
\includegraphics[width=3.5in]{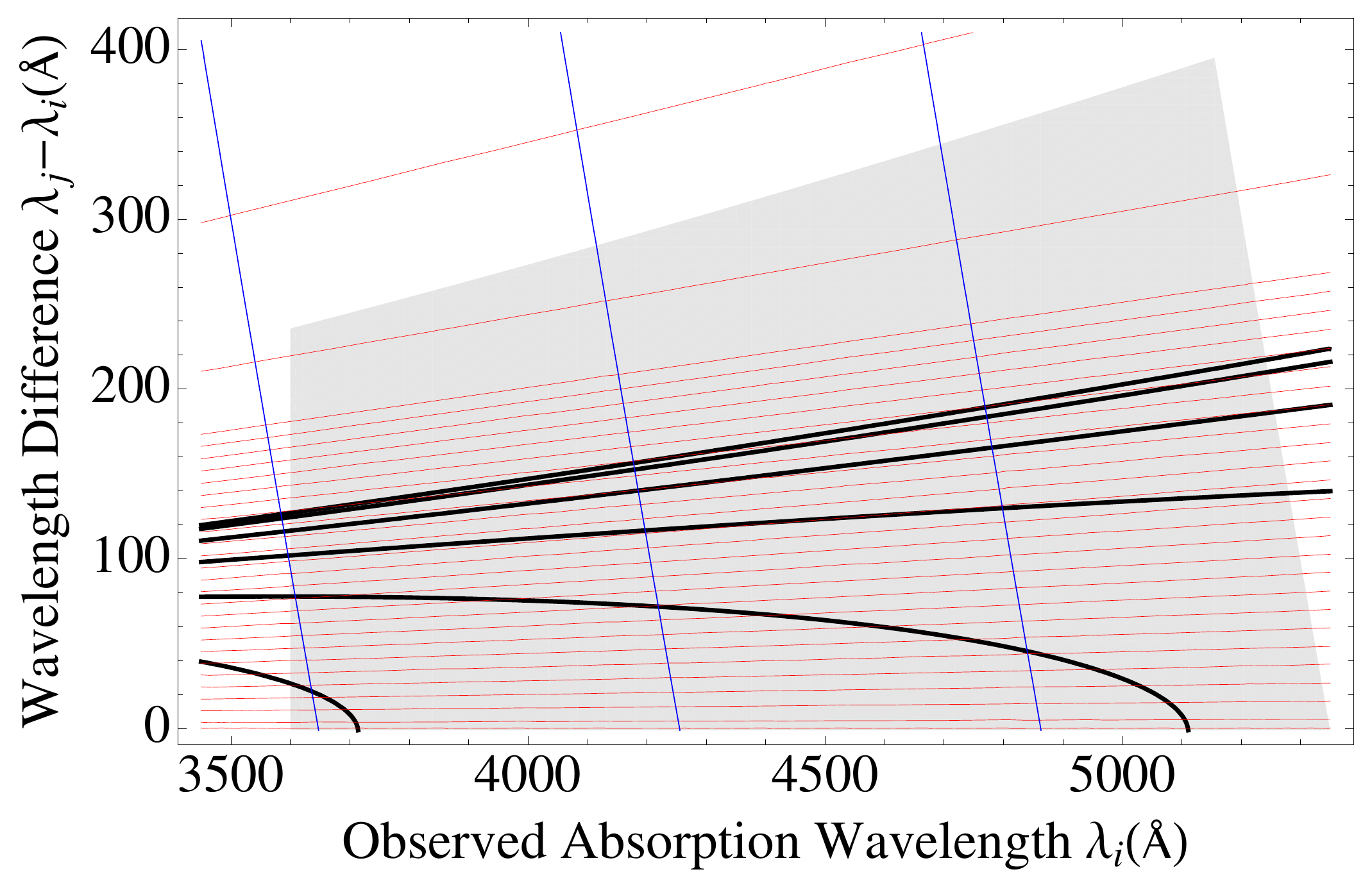}
\caption{Physical coordinates for pixel pairs with observed Lyman-$\alpha$ absorption wavelengths $\lambda_i \le \lambda_j$. Grid lines of $z_{ij}$ (vertical blue, values left to right are 2.25, 2.75, 3.25), $\Delta v_{ij}/c$ (horizontal red, values bottom to top cover 0.001--0.049 with 0.002 spacing, with additional contours at 0, 0.059, and 0.083) represent the nominal sampling grid used in a fit. The shaded gray region shows the pixel pairs contributing to a typical BAO fit, bounded by $\lambda_1 > 3600$~\AA\ , $r_{\parallel} < 170$ Mpc/h, and $z_{ij} < 3.25$. Contours of $\Delta\theta$ (thick black curves) at which the 3D separation is 110 Mpc/h (values from bottom left corner out are 100, 80, 60, 40, 20, 0 arcmins) identify pixel pairs contributing to the BAO peak region at different angular separations.}
\label{fig:coords}
\end{center}
\end{figure}

The physical coordinates for a pair of pixels $(i,j)$ measured in the absorption spectra of two quasars are the separation angle $\Delta\theta_{ij}$ between the quasar lines of sight ($\Delta\theta_{ij} = 0$ if the pixels are taken from the same quasar's spectrum) and the observed absorption wavelengths $\lambda_i$ and $\lambda_j$. We convert the observed wavelengths to a relative velocity for the absorption systems\footnote{This definition is not identical to the Doppler velocity that an observer at one absorber would measure for the other absorber, but does agree to second order in the wavelength ratio.}~\cite{2006ApJS..163...80M}
\begin{equation}
\Delta v_{ij} = c\,\log\left( \lambda_j / \lambda_i \right)
\end{equation}
and an average absorption redshift
\begin{equation}
z_{ij} = \frac{\sqrt{\lambda_i \lambda_j}}{\lambda_{\alpha}} - 1
\end{equation}
where $\lambda_{\alpha} \simeq 1216$~\AA\  is the rest wavelength of the Lyman-$\alpha$ transition which determines pixel redshifts $z_i = \lambda_i/\lambda_{\alpha} - 1$. We calculate the corresponding co-moving separations along ($r_{\parallel}$) and perpendicular to ($r_{\perp}$) the line of sight as
\begin{align}
r_{\parallel} &= D_{C,\text{fid}}(z_j) - D_{C,\text{fid}}(z_i) = \frac{1+z_{ij}}{H_{\text{fid}}(z_{ij})}\cdot \Delta v_{ij}\left[ 1 + {\cal O}\left(\Delta v_{ij}/c\right)^2 \right]
\label{eqn:rpar-defn}\\
r_{\perp} &= D_{A,\text{fid}}(z_{ij})\cdot \Delta\theta_{ij} = c \int_0^{z_{ij}} \frac{dz'}{H_{\text{fid}}(z')}\cdot \Delta\theta_{ij}
\label{eqn:rper-defn}
\end{align}
where $D_{C,\text{fid}}(z)$ and $D_{A,\text{fid}}(z)$ are the co-moving line of sight and angular distance functions, respectively, for the assumed fidicual cosmology $H_{\text{fid}}(z)$. We introduce separate scale factors $\alpha_{\parallel}$ and $\alpha_{\perp}$ in Section~\ref{sec:scale-factors} to allow for small discrepancies between the true cosmology and our assumed fiducial model. Fig~\ref{fig:coords} shows that the BOSS blue camera wavelength limit of $\lambda \gtrsim 3600$~\AA\  limits pixel pairs contributing to a BAO peak feature near 110~Mpc/h to $\Delta\theta_{ij} \lesssim 100$ arcmin. Similarly, the BOSS redshift coverage $z_{ij} \lesssim 3.25$ limits the BAO peak to $\Delta v_{ij}/c = \log(\lambda_2/\lambda_1) \lesssim 0.04$ and observed wavelength differences $|\lambda_j - \lambda_i| \lesssim 200$~\AA.

\subsection{Cosmological Models}

We build the cosmological model starting from an isotropic linear power spectrum prediction $\tilde{P}(k,z_0)$ at some reference redshift $z_0$, then embed this prediction in redshift space (we use tildes to denote linear-theory predictions without any redshift space distortions). In the general case of a plane-parallel redshift-space distortion $r \rightarrow (r,\mu)$ we have~\cite{1998ASSL..231..185H}:
\begin{equation}
\xi_{\text{cosmo}}(r,\mu,z_0) = \sum_{\ell\,\text{even}}\, L_{\ell}(\mu)\,\xi_{\ell,\text{cosmo}}(r, z_0)
\label{eqn:xi-multipoles-general}
\end{equation}
with
\begin{equation}
\xi_{\ell,\text{cosmo}}(r,z_0) = \frac{i^{\ell}}{2\pi^2}\, \int_0^{\infty}\, k^2 j_{\ell}(k r) \,P_{\ell}(k,z_0)\,dk
\label{eqn:xil-defn}
\end{equation}
where $L_{\ell}$ is the Legendre polynomial, $j_{\ell}$ is the spherical Bessel function, and $P_{\ell}(k,z_0)$ are the multipoles of the redshift-distorted power spectrum $P(k,\mu_k,z_0)$ with $\mu_k \equiv \hat{z}\cdot\hat{k}$:
\begin{equation}
P_{\ell}(k,z_0) = \frac{2\ell + 1}{2}\, \int_{-1}^{+1} P(k,\mu_k,z_0)\, L_{\ell}(\mu_k)\,d\mu_k \; .
\label{eqn:pl-defn}
\end{equation}
Specializing to linear theory and the distant observer approximation~\cite{1987MNRAS.227....1K}, the infinite series of eqn.~(\ref{eqn:xi-multipoles-general}) is truncated at $\ell = 4$, with
\begin{equation}
\tilde{P}_{\ell}(k,z_0) = b^2(z_0) C_{\ell}(\beta(z_0)) \tilde{P}(k,z_0)
\label{eqn:Ptilde}
\end{equation}
and
\begin{equation}
C_{\ell}(\beta) \equiv \frac{2\ell+1}{2}\,\int_{-1}^{+1}\,\left(1 + \beta\mu_k^2\right)^2 L_{\ell}(\mu_k)\,d\mu_k =
\begin{cases}
1 + \frac{2}{3}\beta + \frac{1}{5}\beta^2 & \ell = 0 \\
\frac{4}{3}\beta + \frac{4}{7}\beta^2 & \ell = 2 \\
\frac{8}{35}\beta^2 & \ell = 4
\end{cases} \; ,
\label{eqn:Cell-defn}
\end{equation}
where $b(z)$ and $\beta(z)$ are the tracer bias and redshift-space distortion parameter at redshift $z$, respectively. We can therefore write
\begin{equation}
\xi_{\text{cosmo}}(r,\mu,z_0) = b^2(z_0)\, \sum_{\ell=0,2,4}\, C_{\ell}(\beta(z_0)) L_{\ell}(\mu)\,\tilde{\xi}_{\ell,\text{cosmo}}(r, z_0)
\label{eqn:xi-expansion}
\end{equation}
in terms of the undistorted linear-theory multipoles
\begin{equation}
\tilde{\xi}_{\ell,\text{cosmo}}(r,z_0) = \frac{i^{\ell}}{2\pi^2}\, \int_0^{\infty}\, k^2 j_{\ell}(k r)\,\tilde{P}(k,z_0)\,dk \; .
\label{eqn:xi-multipoles}
\end{equation}

\begin{figure}[htb]
\begin{center}
\includegraphics[width=6in]{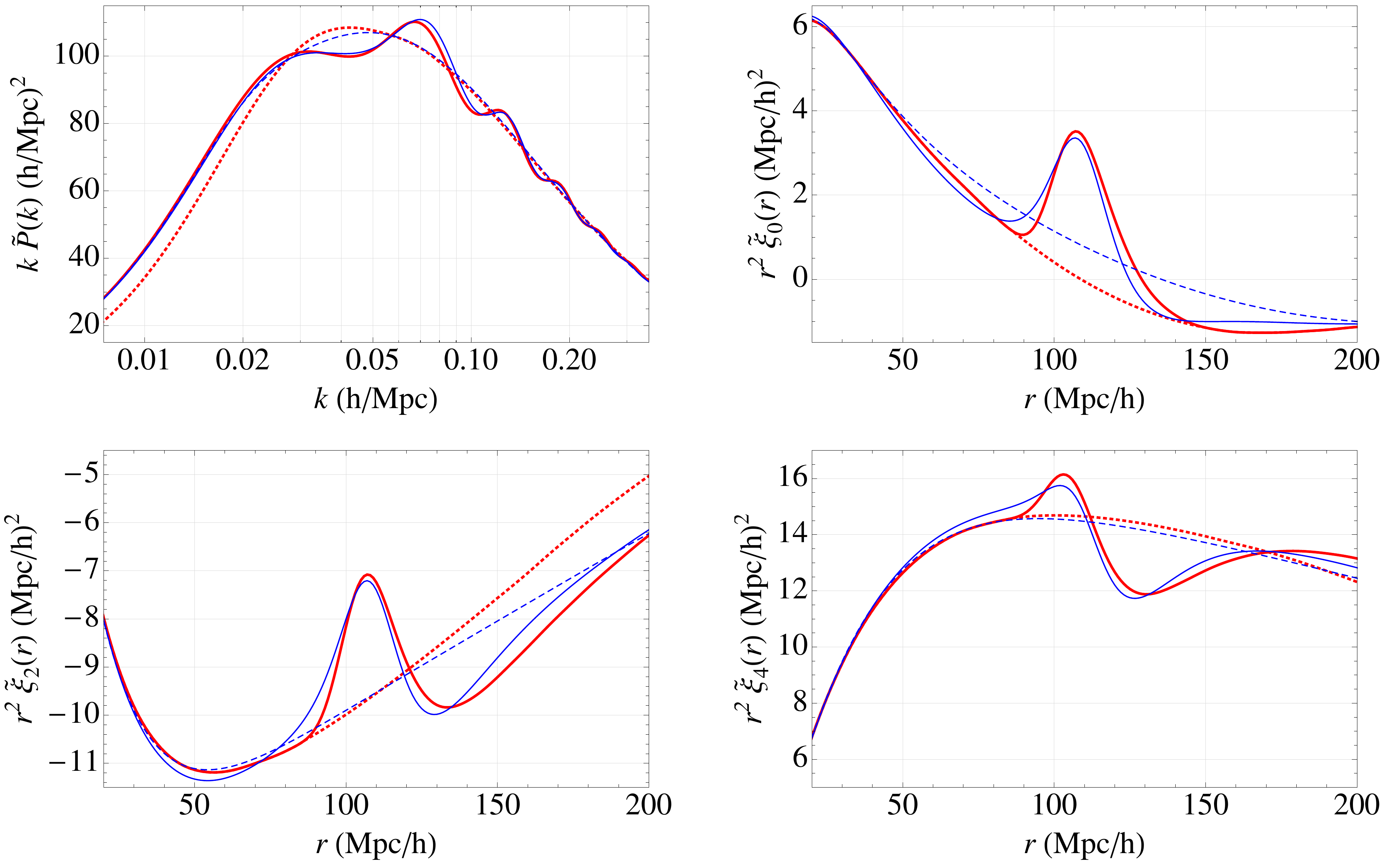}
\caption{Cosmological linear models calculated for $z_0 = 2.25$ and assuming a flat universe with $\Omega_\Lambda = 0.73$, $h = 0.7$, $\Omega_{\text{b}} h^2 = 0.0227$, and $n_s = 0.97$. Panels show the ($k$-weighted) power spectrum (top-left) and the ($r^2$-weighted) correlation function monopole (top-right), quadrupole (bottom-left), and hexadecapole (bottom-right). Curves are calculated with CAMB~\cite{2000ApJ...538..473L} (thick,red) and using ref.~\cite{1998ApJ...496..605E} (light,blue) with solid curves showing the full cosmological model and dotted (dashed) curves showing the corresponding CAMB ``sideband'' (``no-wiggles'' of ref.~\cite{1998ApJ...496..605E}) smooth model.}
\label{fig:cosmo-models}
\end{center}
\end{figure}

Figure~\ref{fig:cosmo-models} shows examples of the linear models we use for Lyman-$\alpha$ fitting in this paper, calculated with $z_0 = 2.25$. Note that the undistorted multipoles $\tilde{\xi}_{\ell}$ are not independent since they derive from the same underlying power spectrum via eqn.~(\ref{eqn:xi-multipoles}). Explicitly, we find that $\tilde{\xi}_2(r)$ and $\tilde{\xi}_4(r)$ can be calculated directly from $\tilde{\xi}_0(r')$ specified on an interval $r_0 \le r' \le r$ via (we have dropped $z_0$ here for clarity):
\begin{equation}
\begin{aligned}
\tilde{\xi}_2(r) &= \tilde{\xi}_0(r) + \left(\frac{r_0}{r}\right)^3 \left[ \tilde{\xi}_2(r_0) - \tilde{\xi}_0(r_0) \right]
-\frac{3}{r^3} \int_{r_0}^r \tilde{\xi}_0(r') r'^2 dr' \\
\tilde{\xi}_4(r) &= \tilde{\xi}_0(r) + \left(\frac{r_0}{r}\right)^5 \left[ \tilde{\xi}_4(r_0) - \tilde{\xi}_0(r_0) \right]
-\frac{5}{r^5} \int_{r_0}^r \left[ \tilde{\xi}_0(r') + \tilde{\xi}_2(r') \right] r'^4 dr' \; ,
\end{aligned}
\label{eqn:multipoles}
\end{equation}
where $r_0$ is an arbitrarily chosen scale and the influence of scales beyond $[r_0,r]$ is fully specified by the constants of integration $\tilde{\xi}_{\ell}(r_0)$.

\subsubsection{Peak Decomposition}
\label{sec:peak-decomposition}

We expect a Lyman-$\alpha$ analysis to distort the measured broadband shape of the correlation function, so our goal is to only use information from the localized peak near $r \simeq 110$ Mpc/h to measure BAO parameters. To achieve this goal, it is useful to decompose $\xi_{\text{cosmo}}$ into separate ``peak'' and ``smooth'' (or ``no-wiggles'') components. Following ref.~\cite{1998ApJ...496..605E}, we can isolate the oscillations in a multiplicative term of the effective baryon transfer function $T_b(k)$ (see equation (16) in ref.~\cite{1998ApJ...496..605E}), which leads to a rather complicated decomposition for the correlation function. Instead, we adopt the unphysical but more tractable decomposition
\begin{equation}
\tilde{\xi}_{\ell,\text{cosmo}}(r,z_0) = \tilde{\xi}_{\ell,\text{smooth}}(r,z_0) + \tilde{\xi}_{\ell,\text{peak}}(r,z_0)
\end{equation}
with the understanding that this model is only valid when using parameter values that are sufficiently close to their nominal values to recover a physically plausible $\tilde{\xi}_{\ell,\text{cosmo}}$. 

Figure~\ref{fig:cosmo-models} shows the smooth $\tilde{P}(k,z_0)$ and corresponding correlation multipoles suggested in ref.~\cite{1998ApJ...496..605E}: the oscillations are effectively removed, but the corresponding ``peak'' feature, defined as the difference between the full and ``no-wiggles'' models, is not well localized in any of the multipoles, with deviations from the dashed curves extending far from the peak. To remedy this problem, we construct an alternate peak model that is explicitly localized, but only for a single linear combination of multipoles (or, equivalently, a single value of $\mu$) because the integrals in eqn.~(\ref{eqn:multipoles}) effectively spread the peak to all scales above and/or below (depending on the choice of $r_0$) for other linear combinations. Our CAMB ``sideband'' smooth model is constructed as follows: we first isolate a localized peak in the CAMB prediction~\cite{2000ApJ...538..473L} of Figure~\ref{fig:cosmo-models} by simultaneously fitting the regions 50--86 and 150--190 Mpc/h of the monopole to the form
\begin{equation}
\xi_{0,\text{fit}}(r,z_0) = \sum_{j=-3}^{+1}\, c_j r^j \; .
\end{equation}
Next, we replace $\tilde{\xi}_{0,\text{cosmo}}$ with $\xi_{0,\text{fit}}$ in the region 86--150 Mpc/h to obtain $\tilde{\xi}_{0,\text{smooth}}$. We then calculate $\tilde{\xi}_{2,\text{peak}}$ and $\tilde{\xi}_{4,\text{peak}}$ using eqn.~(\ref{eqn:multipoles}) with $\tilde{\xi}_{\ell,\text{peak}}(r_0) = 0$ at $r_0 = 0$. Finally, we calculate
\begin{equation}
\tilde{\xi}_{\ell,\text{smooth}}(r,z_0) = \tilde{\xi}_{\ell,\text{cosmo}}(r,z_0) - \tilde{\xi}_{\ell,\text{peak}}(r,z_0) \; .
\end{equation}
The resulting CAMB ``sideband'' smooth model is shown in Figure~\ref{fig:cosmo-models}.

\begin{figure}[htb]
\begin{center}
\includegraphics[width=6in]{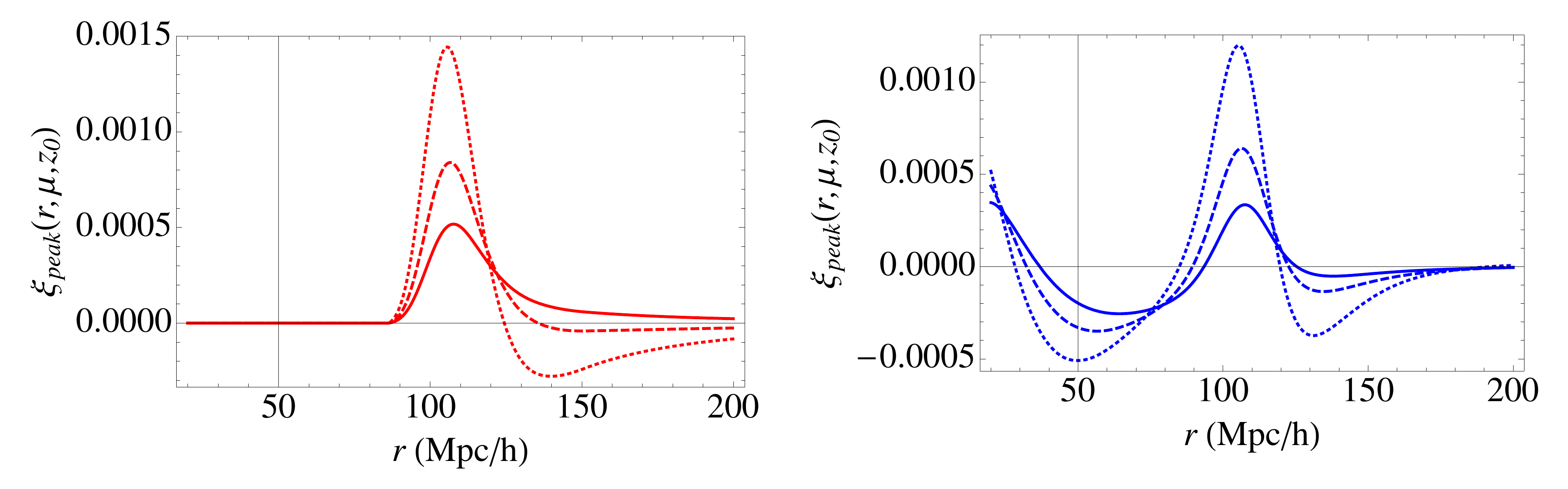}
\caption{Cosmological peak models calculated using the CAMB ``sideband'' method described in the text (left, red) and the ``no-wiggles'' method of ref.~\cite{1998ApJ...496..605E} (right, blue) described in the text. Curves show $\mu =$ 0.4 (solid), 0.7 (dashed), and 1.0 (dotted). There is no $r^2$ weighting applied here.}
\label{fig:mu-slices}
\end{center}
\end{figure}

\begin{figure}[htb]
\begin{center}
\includegraphics[width=6in]{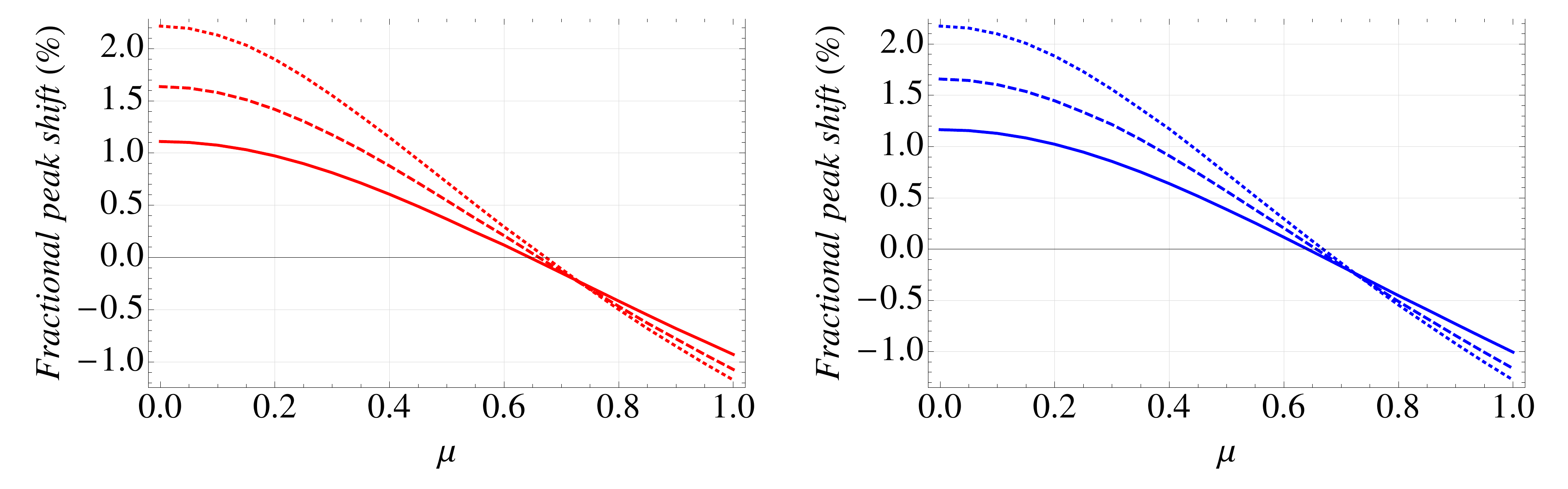}
\caption{Evolution of the peak position with $\mu$ for the cosmological peak models shown in Figure~\ref{fig:mu-slices}. Curves show $\beta =$ 1.0 (solid), 1.4 (dashed), and 1.8 (dotted). Fractional shifts are measured relative to the position of the monopole peak for each model.}
\label{fig:peak-shift}
\end{center}
\end{figure}

By construction, our CAMB peak multipoles are exactly zero below 86 Mpc/h, but the peak spreads to large scales for $\ell = 2,4$, as required by eqn.~(\ref{eqn:multipoles}). Note that although it would be possible to specify independent localized peaks for each multipole, this is unphysical (within the framework of ref.~\cite{1987MNRAS.227....1K}) and does not, in general, reproduce the evolution of the peak shape in $\xi(r,\mu,z_0)$ with $\mu$ implied by the constraint of a single underlying $\tilde{P}(k,z_0)$, as illustrated in Figure~\ref{fig:mu-slices}. For example, the constrained evolution leads to a percent-level shift in the peak position as a function of $\mu$ shown in Figure~\ref{fig:peak-shift} for the two models used here. Note that this shift is essentially the same for both models, despite the rather different peak shapes shown in Figure~\ref{fig:mu-slices}.

\subsubsection{Nonlinear Effects}
\label{sec:nonlinear-effects}

The expected effects of non-linear structure growth on the BAO feature can be modeled with an anisotropic Gaussian roll-off of the linear power spectrum~\cite{2007ApJ...664..660E}:
\begin{equation}
\tilde{P}_{\text{NL}}(k,\mu_k,z_0) = \exp(-k^2 \Sigma^2(\mu_k)/2)\cdot \tilde{P}(k,z_0)
\label{eqn:nl-broadening-defn}
\end{equation}
where
\begin{equation}
\Sigma^2(\mu_k) = \mu_k^2 \Sigma_{\parallel}^2 + (1-\mu_k^2) \Sigma_{\perp}^2 \; .
\end{equation}
In general, this approach breaks the decomposition, eqn.~(\ref{eqn:Ptilde}), of $\tilde{P}_{\ell}(k,z_0)$ into separate $\beta$- and $k$-dependent factors and requires that the integrals of eqns.~(\ref{eqn:xil-defn})--(\ref{eqn:pl-defn}) be re-evaluated for each value of $\beta$. However, since we expect $\Sigma \simeq 5$ Mpc/h, compared with an expected peak full-width half-maximum of $\simeq 25$ Mpc/h, we can approximate for $\beta \simeq \beta_0$:
\begin{equation}
P_{\ell,NL}(k,z_0) \simeq \exp(-k^2\Sigma_{\ell}^2(\beta_0)/2) \cdot \tilde{P}_{\ell}(k,z_0)
\end{equation}
with
\begin{equation}
\Sigma^2_{\ell}(\beta) \equiv f_{\ell}(\beta)\cdot\Sigma_{\parallel}^2 + (1 - f_{\ell}(\beta))\cdot\Sigma_{\perp}^2
\end{equation}
and
\begin{equation}
f_{\ell}(\beta) \equiv \frac{\int_{-1}^{+1}\,\mu_k^2 \left(1 + \beta\mu_k^2\right)^2 L_{\ell}(\mu_k)\,d\mu_k}
{\int_{-1}^{+1}\,\left(1 + \beta\mu_k^2\right)^2 L_{\ell}(\mu_k)\,d\mu_k} =
\begin{cases}
\frac{35+42\beta+15\beta^2}{105+70\beta+21\beta^2} & \ell = 0 \\
\frac{7+12\beta+5\beta^2}{14\beta+6\beta^2} & \ell = 2 \\
\frac{15}{11}+\frac{2}{\beta} & \ell = 4
\end{cases} \; .
\end{equation}
This approximation effectively models anisotropic broadening using different amounts of isotropic broadening for each multipole. The resulting correlation function multipoles $\xi_{\ell,NL}$, calculated with eqn.~(\ref{eqn:xil-defn}), can then be substituted in eqn.~(\ref{eqn:xi-expansion}).

\begin{figure}[htb]
\begin{center}
\includegraphics[width=6in]{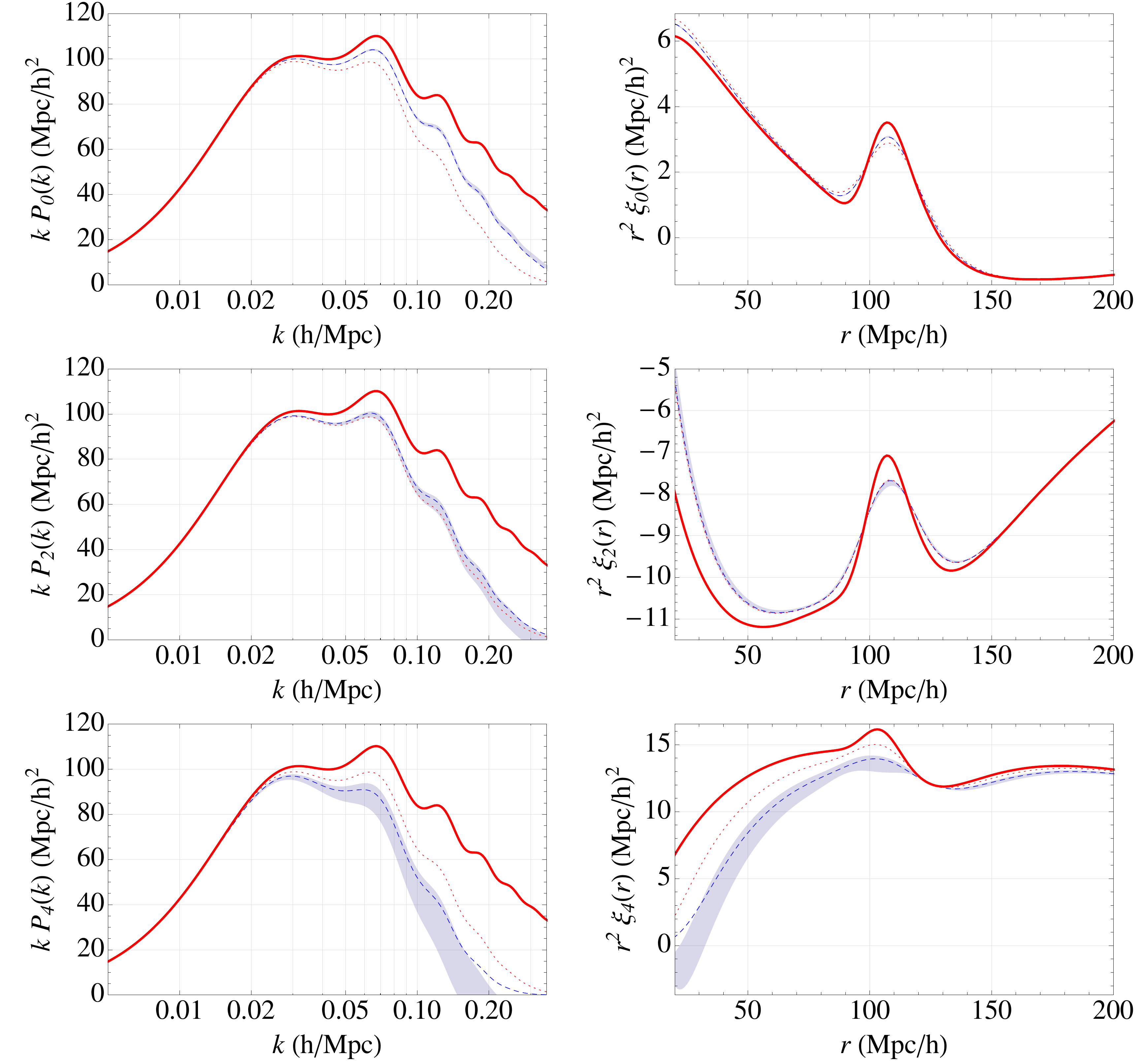}
\caption{Effects of anisotropic non-linear broadening implemented with eqn.~(\ref{eqn:nl-broadening-defn}) and applied to $z_0 = 2.25$ linear CAMB predictions~\cite{2000ApJ...538..473L} using $\Sigma_{\parallel} = 6.41$ Mpc/h and $\Sigma_{\perp} = 3.26$ Mpc/h. Curves show no broadening (thick red, same as curves in Figure~\ref{fig:cosmo-models}), isotropic broadening (dotted red) by $(\Sigma_{\parallel}^2+\Sigma_{\perp}^2)^{1/2}/2 = 5.09$ Mpc/h, the approximate anisotropic model described in the text (dashed blue) with $\beta_0 = 1.4$, and the envelope of full anisotropic calculations (light blue shaded) for $\beta =$ 0.5--2.5. Left-hand panels show the $k$-weighted multipoles $P_{\ell,NL}(k,z_0)$ with $b^2 C_{\ell}(\beta)$ divided out, for $\ell = 0$ (top), $\ell = 2$ (middle), and $\ell = 4$ (bottom). Right-hand panels show the corresponding $r^2$-weighted correlation function multipoles $\xi_{\ell}(r,z_0)$.}
\label{fig:cosmo-nonlinear-models}
\end{center}
\end{figure}

Taking a fiducial value of $\beta_0 = 1.4$, we calculate $f_0 = 0.505$, $f_2 = 1.07$, and $f_4 = 2.79$. At a redshift $z = 2.4$, we expect $\Sigma_{\parallel} \simeq 6.41$ Mpc/h and $\Sigma_{\perp} \simeq 3.26$ Mpc/h, so that $\Sigma_0 = 5.10$ Mpc/h, $\Sigma_2 = 6.58$ Mpc/h, and $\Sigma_4 = 9.79$ Mpc/h. Figure~\ref{fig:cosmo-nonlinear-models} compares the resulting approximate models with exact calculations for a range of $\beta$ values. Note that we are neglecting the non-zero even multipoles $\ell = 6,8,\ldots$ that are introduced by anisotropic non-linear broadening with this approximation, but Fig~\ref{fig:cosmo-nonlinear-models} shows that most of the BAO peak feature has already been smoothed out by $\ell = 4$. We also neglect any redshift evolution of the non-linear broadening parameters, since this evolution would be a second order effect on what is already a small correction to the linear theory for the purposes of measuring the BAO feature.

\begin{figure}[htb]
\begin{center}
\includegraphics[width=6in]{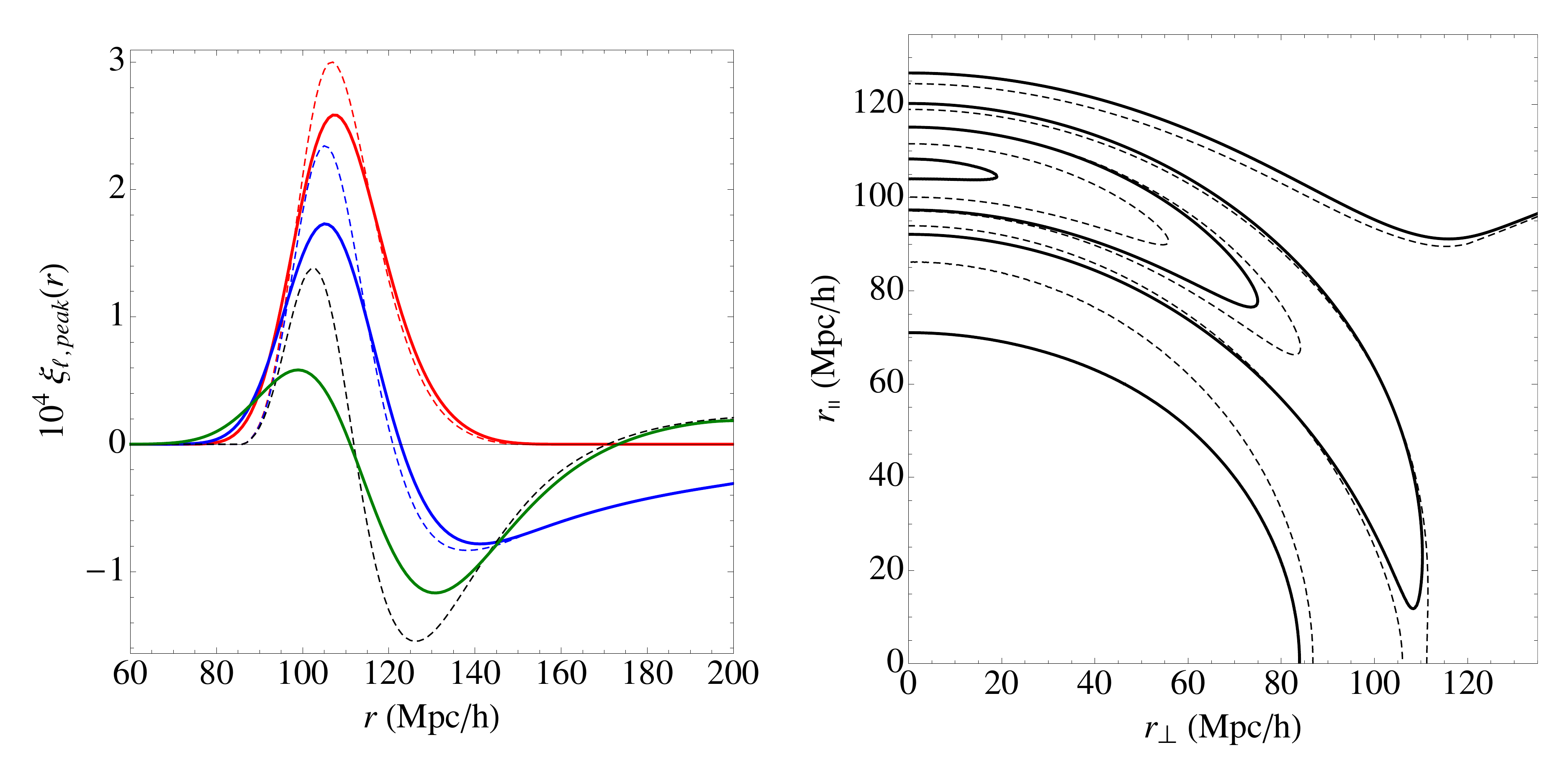}
\caption{Effects of anisotropic non-linear broadening on peak templates, implemented using the approximations described in the text and applied to $z_0 = 2.25$ linear CAMB predictions~\cite{2000ApJ...538..473L}. Solid (dashed) curves show peak templates with (without) non-linear effects. From top to bottom, curves in the left-hand panel show the $\ell = 0$ (red), 2 (blue), and 4 (green) correlation multipoles, with $b^2 C_{\ell}(\beta)$ divided out, without any $r$-weighting and scaled by $10^4$. All templates are identically zero for $r < 60$ Mpc/h, by construction. The right-hand panel shows equally spaced contours of $\xi_{\text{peak}}(r,\mu)$ calculated with $\beta = 1.4$, with the outer contour corresponding to zero correlation.}
\label{fig:nonlinear-peaks}
\end{center}
\end{figure}

\begin{figure}[htb]
\begin{center}
\includegraphics[width=6in]{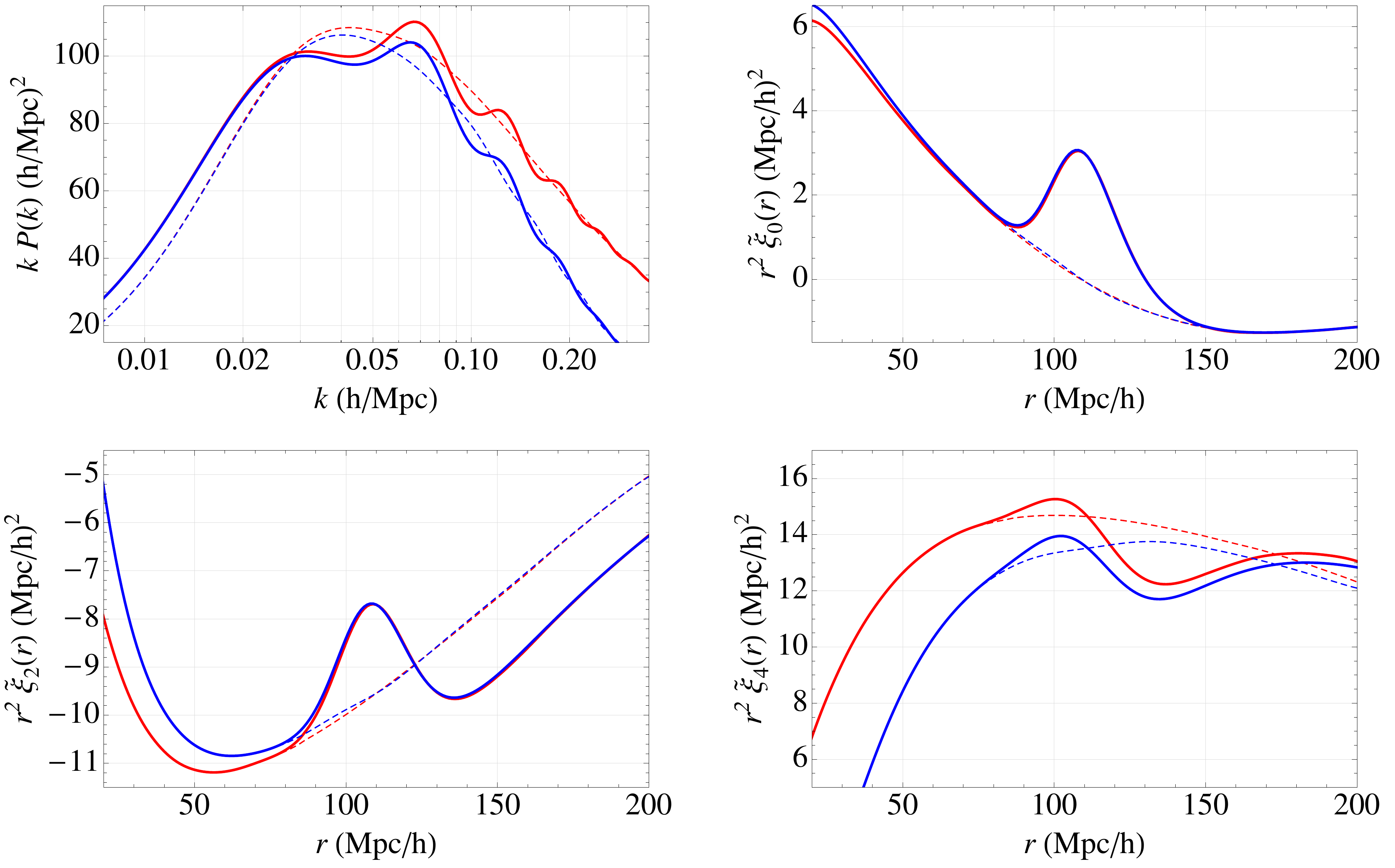}
\caption{Comparison of two different schemes for incorporating anisotropic non-linear effects into the final templates used for fitting. Panels show the ($k$-weighted) power spectrum (top-left) and the ($r^2$-weighted) correlation function monopole (top-right), quadrupole (bottom-left), and hexadecapole (bottom-right). Red curves are calculated with non-linear effects applied to the BAO peak feature only. Blue curves are calculated with non-linear broadening applied at all scales. Both schemes are derived from $z_0 = 2.25$ linear CAMB predictions~\cite{2000ApJ...538..473L} and use the same broadened peaks shown in Figure~\ref{fig:nonlinear-peaks}. The combined peak + smooth (smooth only) templates are represented with solid (dashed) curves.}
\label{fig:nonlinear-model-components}
\end{center}
\end{figure}

In order to achieve a self-consistent decomposition of each correlation-function multipole into peak + smooth components with non-linear broadening effects included, we first broaden each $\xi_{\ell,\text{peak}}(r,z_0)$ by $\Sigma_{\ell}$ to obtain $\xi_{\ell,\text{peak},NL}(r,z_0)$. Broadening by $\Sigma$ in this context implies
\begin{equation}
P(k) \rightarrow P(k) \exp(-k^2 \Sigma^2/2)
\end{equation}
which transforms to
\begin{equation}
\xi_0(r) \rightarrow \int_0^\infty\, ds \,\frac{s}{r}\, \left[ G(r-s,\Sigma) - G(r+s,\Sigma) \right] \xi_0(s) \; ,
\end{equation}
where $G$ is the normalized one-dimensional Gaussian
\begin{equation}
G(t,\Sigma) \equiv \frac{1}{\sqrt{2\pi}\Sigma}\, \exp\left(-\frac{t^2}{2\Sigma^2}\right)\; .
\end{equation}
Note that when $r \gg \Sigma$, we recover the expected convolution
\begin{equation}
\xi_0(r) \rightarrow \int_0^\infty\, ds\, G(r-s,\Sigma)\,\xi_0(s)
\end{equation}
to a good approximation. Figure~\ref{fig:nonlinear-peaks} shows the broadened multipole peaks based on our linear CAMB templates. To derive the corresponding smooth templates, we either use the linear smooth templates, or else we calculate non-linear smooth templates by subtracting the broadened peak templates from the broadened multipoles:
\begin{equation}
\xi_{\ell,\text{smooth},NL} = \xi_{\ell,NL} - \tilde{\xi}_{\ell,\text{peak},NL} \; .
\end{equation}
In the first case, we are only applying non-linear broadening to the peak feature, which introduces an unphysical distinction between peak and smooth components.
In the second case we are broadening the full correlation function, which removes the distinction but applies unphysical filtering of small scale structure. A better model is probably somewhere in between so we consider both alternatives, and expect that any extracted peak parameters will not depend on this choice. Figure~\ref{fig:nonlinear-model-components} compares both approaches.

\subsubsection{Redshift Evolution}
\label{sec:redshift-evolution}

In general, we model the redshift evolution of a parameter $p(z)$ for $z$ near $z_0$ in terms of two parameters $p_0$ and $\gamma_p$ via
\begin{equation}
p(z) = p_0 \, \left(\frac{1+z}{1+z_0}\right)^{\gamma_p} \; .
\end{equation}
We apply this evolution to the parameters $b^2(z)$ and $\beta(z)$, introduced above, and to the BAO scale parameters $\alpha_{\text{iso}}(z)$, $\alpha_{\parallel}(z)$ and $\alpha_{\perp}(z)$ introduced below. Given a covariance matrix for the parameters $p_0$ and $\gamma_p$,
\begin{equation}
C_p =
\begin{pmatrix}
\sigma_0^2 & \rho \sigma_0 \sigma_\gamma \\
\rho \sigma_0 \sigma_\gamma & \sigma_\gamma^2 \\
\end{pmatrix}
\end{equation}
the variance of $p(z)$ with $z \simeq z_0$ is given by
\begin{equation}
\sigma^2_{p(z)} = J\cdot C_p\cdot J^t
\end{equation}
where
\begin{equation}
J = \left( \frac{\partial p(z)}{\partial p_0}, \frac{\partial p(z)}{\partial\gamma_p} \right)
\end{equation}
is the Jacobian. The error on $p(z)$ is smallest at
\begin{equation}
\log \left( \frac{1+z}{1+z_0} \right) = -b - \rho a + \sqrt{b^2 - (1-\rho^2) a^2}
\label{eqn:pz-min-error}
\end{equation}
with
\begin{equation}
a \equiv \frac{\sigma_0}{p_0 \sigma_\gamma} \quad , \quad b \equiv \frac{1}{2\gamma_p} \; .
\end{equation}

\subsubsection{Scale Factors}
\label{sec:scale-factors}

When fitting our model to data, we allow for an overall relative normalization factor $a_{\text{peak}} \simeq 1$ as well as a general coordinate transform, $r'=r'(r,\mu,z)$ and $\mu'=\mu(r,\mu,z)$, that allows for possible small differences\footnote{In case there is evidence for large differences, the analysis should be repeated with a fidicual cosmology that better matches the data.} between the fiducial and actual cosmologies:
\begin{equation}
\xi_{\text{cosmo}}(r,\mu,z) \rightarrow a_{\text{peak}}\cdot \bigl[\xi(r',\mu',z) - \xi_{\text{smooth}}(r',\mu',z)\bigr]
+ \xi_{\text{smooth}}(r'',\mu'',z) \; .
\label{eqn:bband-defn}
\end{equation}
We consider two options for the cosmological broadband $\xi_{\text{smooth}}$: we either apply the same transform as for the peak ($r'' = r'$), or else we keep it fixed ($r'' = r$) so that only the peak is transformed. Neither approach is exact when considering variations of the cosmological parameters around our fidicual model, but we find that decoupling the peak from the cosmological broadband ($r'' = r$) better localizes the separations contributing to a BAO measurement to the peak region (compare Figs.~\ref{fig:fisher-iso} and \ref{fig:fisher-coupled}), so is preferred when broadband distortion is not fully under control. The corresponding $k$-space transforms are defined by
\begin{equation}
k'(k,\mu_k,z)\cdot r'(r,\mu,z) = k\cdot r \quad , \quad \mu_k'(k,\mu_k,z)\cdot \mu(r,\mu,z) = \mu_k\cdot \mu \; .
\end{equation}
Including a $\mu$-dependence in the coordinate transform also enables us to study the constraining power of our data separately along and transverse to the line of sight. Similarly, a $z$-dependence allows us to determine the redshift at which our results are best measured using eqn.~(\ref{eqn:pz-min-error}).

\begin{figure}[htb]
\begin{center}
\includegraphics[width=6in]{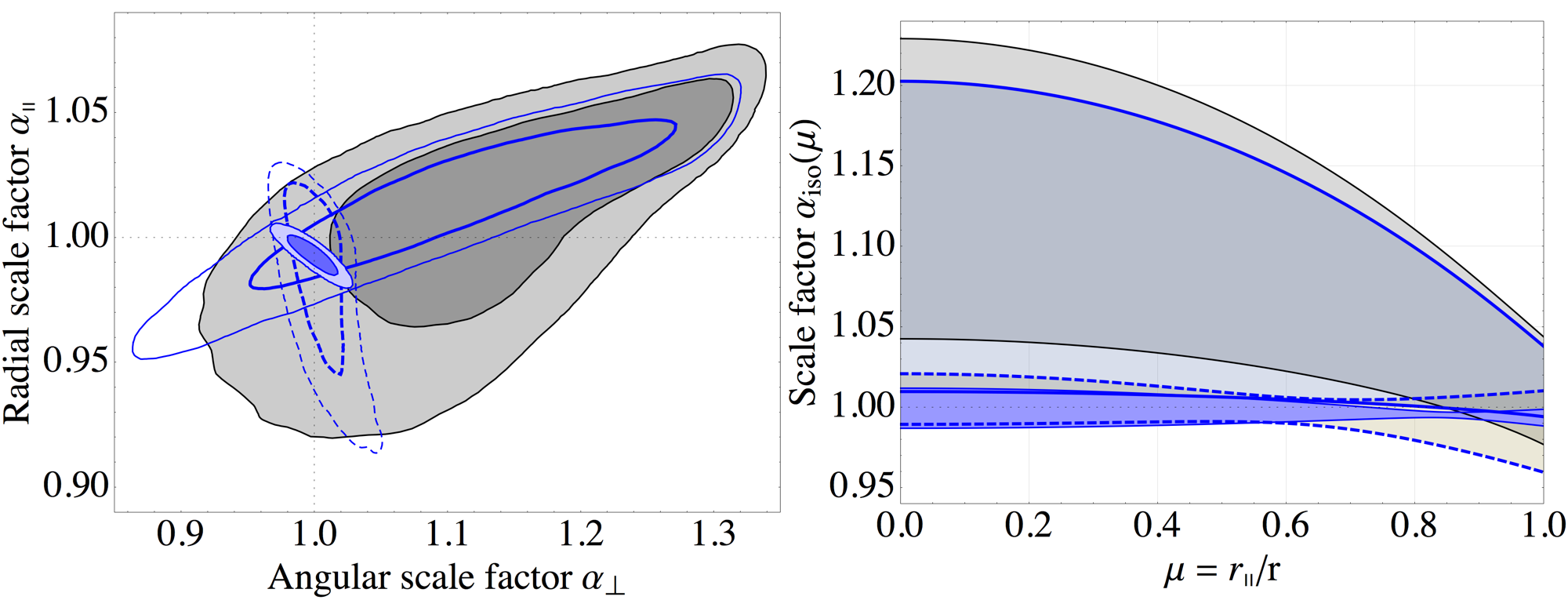}
\caption{Predictions at $z = 2.4$ for the anisotropic scale factors (left) and $\alpha_{\text{ani}}(\mu)$ (right) from WMAP9 data~\cite{2012arXiv1212.5226H}. Filled blue contours show $\Lambda$CDM predictions, while solid (dashed) curves show the effects of allowing the curvature $\Omega_k$ (dark energy equation of state parameter $w_0$) to vary in the model. The gray filled contours show the combined effects of varying both $\Omega_k$ and $w_0$.
Contours in the left-hand plot enclose 68\% and 95\% while the bands in the right-hand plot are $\pm 1$ standard deviation. The fiducial cosmology $\alpha_{\parallel} = \alpha_{\perp} = \alpha(\mu) = 1$ is indicated by dotted lines.}
\label{fig:wmap9-constraints}
\end{center}
\end{figure}

For our baseline isotropic model, we use
\begin{equation}
\begin{aligned}
r'_{\text{iso}}(r,\mu,z) &= \alpha_{\text{iso}}(z)\cdot r \\
\mu'_{\text{iso}}(r,\mu,z) &= \mu \; .
\end{aligned}
\end{equation}
For our baseline anisotropic model, we decouple the line-of-sight ($r_{\parallel}\rightarrow \alpha_{\parallel} r_{\parallel}$) and transverse ($r_{\perp} \rightarrow \alpha_{\perp} r_{\perp}$) scales using
\begin{equation}
\begin{aligned}
r'_{\text{ani}}(r,\mu,z) &= \alpha_{\text{ani}}(\mu,z)\cdot r\\
\mu'_{\text{ani}}(r,\mu,z) &= \frac{\alpha_{\parallel}(z)}{\alpha_{\text{ani}}(\mu,z)}\cdot \mu
\end{aligned}
\end{equation}
with
\begin{equation}
\alpha_{\text{ani}}(\mu,z) = \sqrt{\alpha^2_{\parallel}(z)\mu^2 + \alpha^2_{\perp}(z)(1-\mu^2)} \; .
\end{equation}
When the parameters of this anisotropic model are determined by measuring the physical coordinate separation scales $\Delta v_{\text{BAO}}(z)$  and $\Delta\theta_{\text{BAO}}(z)$ corresponding to the comoving BAO scale $r_{\text{BAO}}$ at some redshift $z$, related by
\begin{align}
\Delta v_{\text{BAO}}(z) &\simeq r_{\text{BAO}}\,H(z)/(1+z) \\
\Delta\theta_{\text{BAO}}(z) &= r_{\text{BAO}}/D_A(z) \; ,
\end{align}
then the expected best fit values of $\alpha_{\parallel}$ and $\alpha_{\perp}$ satisfy
\begin{align}
\alpha_{\parallel}\cdot \frac{1+z}{H_{\text{fid}}(z)} \Delta v_{\text{BAO}}(z) &\simeq r_{\text{BAO,fid}} \\
\alpha_{\perp}\cdot D_{A,\text{fid}}(z) \Delta\theta_{\text{BAO}}(z) &= r_{\text{BAO,fid}}
\end{align}
where $r_{\text{BAO,fid}}(z)$ is the comoving BAO scale predicted by the assumed fiducial cosmology. Combining these results, the best-fit anisotropic scale factors measure
\begin{equation}
\alpha_{\parallel}(z) = \frac{r_{\text{BAO,fid}}}{r_{\text{BAO}}}\cdot \frac{H_{\text{fid}}(z)}{H(z)}
\quad , \quad
\alpha_{\perp}(z) = \frac{r_{\text{BAO,fid}}}{r_{\text{BAO}}}\cdot \frac{D_A(z)}{D_{A,\text{fid}}(z)} \; .
\end{equation}
The distributions of these ratios predicted at $z = 2.4$ by WMAP9~\cite{2011ApJS..192...18K} observations of the CMB are shown in Figure~\ref{fig:wmap9-constraints}.

\begin{figure}[htb]
\begin{center}
\includegraphics[width=6in]{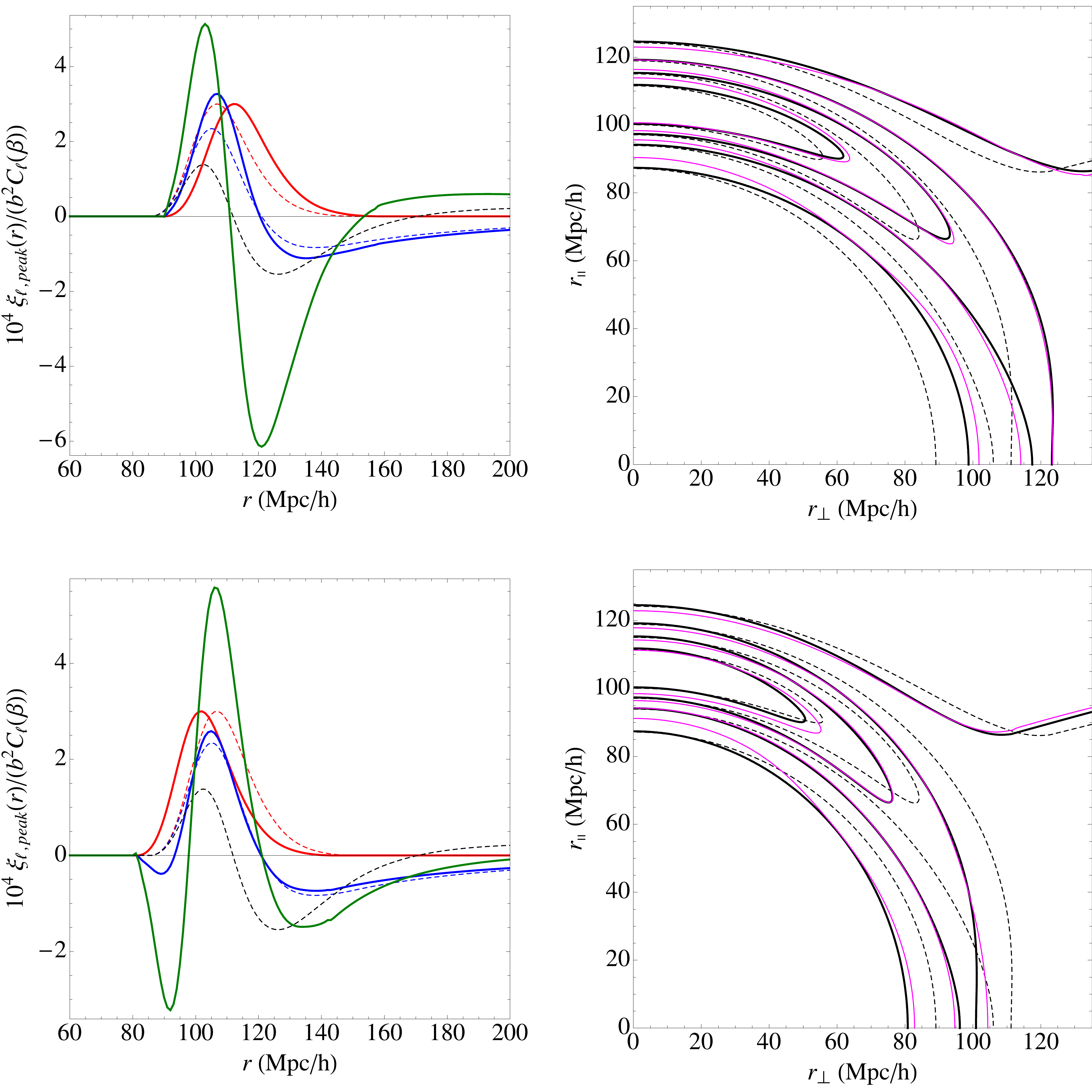}
\caption{Examples of anisotropic coordinate transforms applied to the correlation function, using $\beta = 1.4$. The top row is calculated with $\alpha = 0.95$, $\epsilon = 0.05$ and the bottom row with $\alpha = 1.05$, $\epsilon = -0.05$, both chosen to match the fiducial cosmology at $\mu = 1$, where the expected sensitivity is largest for $\beta = 1.4$. Left-hand plots show the normalized distorted correlation multipoles $10^4 \xi_{\ell}(r)/(b^2 C_{\ell}(\beta))$ calculated with eqn.~(\ref{eqn:scaled-anisotropic-multipoles}) (solid curves) or without any coordinate scaling (dashed curves). From top to bottom, curves show $\ell =$ 0 (red), 2 (blue), and 4 (green). Right-hand plots show equally spaced contours of $\xi(r,\mu,z_0)$ with the outer contours corresponding to zero correlation. Thick solid curves are calculated with eqn.~(\ref{eqn:anisotropic-rmu}) and dashed curves are calculated without any coordinate scaling. Thin magenta curves show the results of combining the transformed multipoles (solid curves) of the left-hand plots using eqn.~(\ref{eqn:xil-defn}), and demonstrate the level of accuracy provided by the first-order approximations described in the text.}
\label{fig:anisotropic-peaks}
\end{center}
\end{figure}

We can evaluate a transformed correlation function $\xi'$ in the original coordinates $(r,\mu,z)$ using eqn.~(\ref{eqn:xi-expansion}) (with $z$-dependencies omitted for clarity)
\begin{equation}
\begin{aligned}
\xi'(r,\mu) &= \xi_{\text{cosmo}}(r'(r,\mu),\mu'(r,\mu))\\
&= b^2 \sum_{\ell = 0,2,4}\, C_{\ell}(\beta)\,L_{\ell}(\mu'(r,\mu))\,
\tilde{\xi}_{\ell,\text{cosmo}}(r'(r,\mu))  \; .
\end{aligned}
\label{eqn:anisotropic-rmu}
\end{equation}
The corresponding multipoles in the original coordinates are then given by 
\begin{align}
\xi'_{\ell}(r) &\equiv \frac{2\ell+1}{2}\, \int_{-1}^{+1}\, \xi'(r,\mu)\,L_{\ell}(\mu) d\mu \\
&= \frac{2\ell+1}{2}\, b^2 \sum_{\ell'=0,2,4} C_{\ell}(\beta)\,\int_{-1}^{+1}\,
\tilde{\xi}_{\ell,\text{cosmo}}(r'(r,\mu))\,L_{\ell}(\mu'(r,\mu))\cdot L_{\ell}(\mu) d\mu \; .
\label{eqn:transformed-multipoles}
\end{align}
In the case of isotropic distortion, we can simply replace $r$ with $\alpha_{\text{iso}}(z)\cdot r$ in the multipoles obtained with eqn.~(\ref{eqn:xi-multipoles}). However, anisotropic distortions mix multipoles and lead to more complicated expressions. For small $\epsilon \equiv (\alpha_{\parallel} - \alpha_{\perp})/(\alpha_{\parallel} + \alpha_{\perp})$ and arbitrary $\alpha \equiv (\alpha_{\parallel} + \alpha_{\perp})/2$, we have
\begin{equation}
\alpha_{\parallel} = \alpha(1+\epsilon) \quad , \quad \alpha_{\perp} = \alpha(1 - \epsilon)
\end{equation}
and corresponding transforms
\begin{equation}
\begin{aligned}
r_{\text{ani}}'(r,\mu) &= \alpha r \left[ 1 + 2\epsilon(2\mu^2-1) + \epsilon^2\right]^{1/2}
&\simeq \alpha r \left[ 1 + (2\mu^2-1)\cdot \epsilon\,\right] \\
\mu_{\text{ani}}'(r,\mu) &= \mu(1+\epsilon) \left[ 1 + 2\epsilon(2\mu^2-1) + \epsilon^2\right]^{-1/2}
&\simeq \mu \left[ 1 + 2(1-\mu^2)\cdot \epsilon\,\right] \; .
\end{aligned}
\end{equation}
Using these approximations, and
\begin{equation}
\tilde{\xi}_{\ell,\text{cosmo}}(r'(r,\mu)) \simeq
\tilde{\xi}_{\ell,\text{cosmo}}(\alpha r) + \epsilon\cdot \alpha r \cdot
\partial_r\tilde{\xi}_{\ell',\text{cosmo}}(\alpha r)\cdot (2\mu^2 - 1) \; ,
\label{eqn:taylor-expansion}
\end{equation}
we obtain distorted multipoles:
\begin{equation}
\xi'_{\ell}(r) \simeq b^2 \sum_{\ell'=0,2,4} C_{\ell'}(\beta)\,\left\{
\tilde{\xi}_{\ell',\text{cosmo}}(\alpha r)\cdot A_{\ell,\ell'}(\epsilon) +
\epsilon \cdot \alpha r \cdot\partial_r\tilde{\xi}_{\ell',\text{cosmo}}(\alpha r)\cdot B_{\ell,\ell'}
\right\}
\label{eqn:scaled-anisotropic-multipoles}
\end{equation}
where
\begin{align}
A_{\ell,\ell'}(\epsilon) &\equiv \frac{2\ell+1}{2}\,\int_{-1}^{+1}\,
L_{\ell'}(\mu + 2 \epsilon\mu(1-\mu^2)) L_{\ell}(\mu) d\mu \\
B_{\ell,\ell'} &\equiv \frac{2\ell+1}{2}\,\int_{-1}^{+1}\,(2\mu^2-1)
L_{\ell'}(\mu) L_{\ell}(\mu) d\mu \; .
\end{align}
We find the following non-zero coefficients, to first order in $\epsilon$:
\begin{gather}
A_{0,0} = 1 \quad , \quad B_{0,0} = -\frac{1}{3} \quad , \quad
A_{0,2} = \frac{4}{5}\epsilon \quad , \quad B_{0,2} = \frac{4}{15} \\
B_{2,0} = \frac{4}{3} \quad , \quad
A_{2,2} = 1 + \frac{4}{7}\epsilon \quad , \quad B_{2,2} = \frac{1}{21} \quad , \quad
A_{2,4} = \frac{40}{21}\epsilon \quad , \quad B_{2,4} = \frac{8}{21} \\
A_{4,2} = -\frac{48}{35}\epsilon \quad , \quad B_{4,2} = \frac{24}{35} \quad , \quad
A_{4,4} = 1 + \frac{40}{77}\epsilon \quad , \quad B_{4,4} = \frac{1}{77} \\
A_{6,4} = -\frac{80}{33}\epsilon \quad , \quad B_{6,4} = \frac{20}{33} \; .
\end{gather}
Note that an $\ell = 6$ term must be included in order to match eqn.~(\ref{eqn:transformed-multipoles}) to first order in $\epsilon$, but is numerically negligible. On the other hand, second-derivative terms in the Taylor expansion~\ref{eqn:taylor-expansion} are formally ${\cal O}(\epsilon^2)$ but are numerically significant when
\begin{equation}
\epsilon \gtrsim \frac{2 |\partial_r\tilde{\xi}_{\ell}(\alpha r)|}{\alpha r |\partial^2_r\tilde{\xi}_{\ell}(\alpha r)|} \; ,
\end{equation}
which is guaranteed to occur at the peak of the BAO feature and at the transition to a rising smooth broadband below the peak.
Figure~\ref{fig:anisotropic-peaks} shows examples of anisotropic coordinate transforms based on eqns.~(\ref{eqn:anisotropic-rmu}) and (\ref{eqn:scaled-anisotropic-multipoles}). Note that eqn.~(\ref{eqn:scaled-anisotropic-multipoles}) is only exact when $\mu^2 = 1/2$ ($r_{\parallel} = r_{\perp}$) and grows less accurate when moving away from the diagonal, while eqn.~(\ref{eqn:anisotropic-rmu}) is exact and therefore preferred when fitting for the anisotropic scale parameters.

\subsection{Broadband Distortion Models}

A Lyman-$\alpha$ analysis measures the correlation function
\begin{equation}
\xi_{ij} \equiv \langle \delta_i \delta_j \rangle - \langle\delta_i\rangle \langle\delta_j\rangle
= \frac{\langle F_i F_j \rangle - \langle F_i \rangle \langle F_j \rangle}{\overline{F}(\lambda_i) \overline{F}(\lambda_j)}
\end{equation}
for the overdensity proxy
\begin{equation}
\delta_i = \frac{F_i - \overline{F}(\lambda_i)}{\overline{F}(\lambda_i)}
\end{equation}
in pixels $i$ measured at rest absorption wavelengths $\lambda_i$, where
\begin{equation}
F_i = \frac{f_i - f_{\text{sky},i}}{f_{\text{cont},i}}
\label{eqn:F-defn}
\end{equation}
is the transmitted flux fraction, relative to the sky flux level $f_{\text{sky},i}$ and normalized to an assumed quasar continuum flux level $f_{\text{cont},i}$, and $\overline{F}(\lambda_i)$ is an assumed mean transmission fraction at $\lambda_i$.

To investigate possible sources of broadband distortion, write
\begin{equation}
\begin{aligned}
f_i &= \tilde{f}_i + \epsilon_i \\
f_{\text{sky},i} &= \tilde{f}_{\text{sky},i} + \epsilon_{\text{sky},i} + s(\lambda_i) \\
f_{\text{cont},i} &= \tilde{f}_{\text{cont},i} + c_i(\mathbf{d})
\end{aligned}
\label{eqn:distort-model}
\end{equation}
where tilde quantities are true values, $\epsilon_i$ and $\epsilon_{\text{sky},i}$ are (zero-mean) noise sources, $s(\lambda_i)$ accounts for any wavelength-dependent residual sky-subtraction bias (as observed in DR9), and $c_i(\mathbf{d})$ describes the continuum modeling error that, in general, depends on the full vector of pixel measurements
\begin{equation}
\mathbf{d} = \left\{ \lambda_j, f_j \right\}_{i=1}^{N_{\text{pixels}}} \; .
\end{equation}
Combining eqns.~(\ref{eqn:F-defn}) and (\ref{eqn:distort-model}), we find:
\begin{equation}
\begin{aligned}
F_i &= \frac{\tilde{f}_i - \tilde{f}_{\text{sky},i} - s(\lambda_i) + \epsilon_i - \epsilon_{\text{sky},i} }
{\tilde{f}_{\text{cont},i} + c_i(\mathbf{d})}\\
&\simeq \tilde{F}_i \left[ 1 - S_i - C_i + E_i \right]
\end{aligned}
\end{equation}
with
\begin{equation}
\tilde{F}_i \equiv \frac{\tilde{f}_i - \tilde{f}_{\text{sky},i}}{\tilde{f}_{\text{cont},i}} \quad , \quad
S_i \equiv \frac{s(\lambda_i)}{\tilde{f}_i - \tilde{f}_{\text{sky},i}} \quad , \quad
C_i \equiv \frac{c_i(\mathbf{d})}{\tilde{f}_{\text{cont},i}} \quad , \quad
E_i \equiv \frac{\epsilon_i - \epsilon_{\text{sky},i}}{\tilde{f}_i - \tilde{f}_{\text{sky},i}} \quad ,
\end{equation}
where we have assumed that $C_i \ll 1$. Averaging over noise and cosmic realizations, we find:
\begin{equation}
\xi_{ij} \simeq r_{ij} \left[ \tilde{\xi}_{ij} \left( 1 + A_{ij} \right) + B_{ij} \right]
\label{eqn:distortion}
\end{equation}
where
\begin{equation}
\tilde{\xi}_{ij} \equiv \frac{\langle\tilde{F}_i\tilde{F}_j\rangle - \langle\tilde{F}_i\rangle \langle\tilde{F}_j\rangle}
{\langle\tilde{F}_i\rangle \langle\tilde{F}_j\rangle}
\end{equation}
is the undistorted cosmological correlation function we seek to measure, $r_{ij}\simeq 1$ describes the effects of any discrepancy between the assumed and true mean transmitted flux fraction via
\begin{equation}
r_{ij} \equiv \frac{\langle\tilde{F}_i\rangle \langle\tilde{F}_j\rangle}{\overline{F}(\lambda_i) \overline{F}(\lambda_j)} \; ,
\end{equation}
$A_{ij}$ is a multiplicative distortion of the true correlation function
\begin{equation}
A_{ij} \equiv \langle S_i S_j\rangle + \langle C_i C_j\rangle + \langle E_i E_j\rangle
+ \langle S_i\rangle\langle C_j\rangle + \langle S_j\rangle\langle C_i\rangle
- \langle S_i\rangle - \langle S_j\rangle - \langle C_i\rangle - \langle C_j\rangle \; ,
\end{equation}
and $B_{ij}$ is an additive distortion
\begin{equation}
B_{ij} \equiv \langle S_i S_j\rangle - \langle S_i\rangle \langle S_j\rangle
+ \langle C_i C_j\rangle - \langle C_i\rangle \langle C_j\rangle + \langle E_i E_j\rangle \; ,
\end{equation}
where we have used $\langle E_i\rangle = 0$ and assumed that $S$, $C$, and $E$ are mutually uncorrelated.

Equation~(\ref{eqn:distortion}) demonstrates that, in general, the measured correlation function can be systematically distorted by multiplicative and additive effects. The actual correlation function used for fitting is a weighted sum of pixels in bins of $(r,\mu,z)$. Any multiplicative distortion is due to a mismatch between the assumed $\overline{F}(\lambda)$ and the true mean transmitted flux for pixels at the same wavelength $\lambda$, whereas additive distortion is due to correlated continuum fit errors.

\begin{figure}[htb]
\begin{center}
\includegraphics[width=6in]{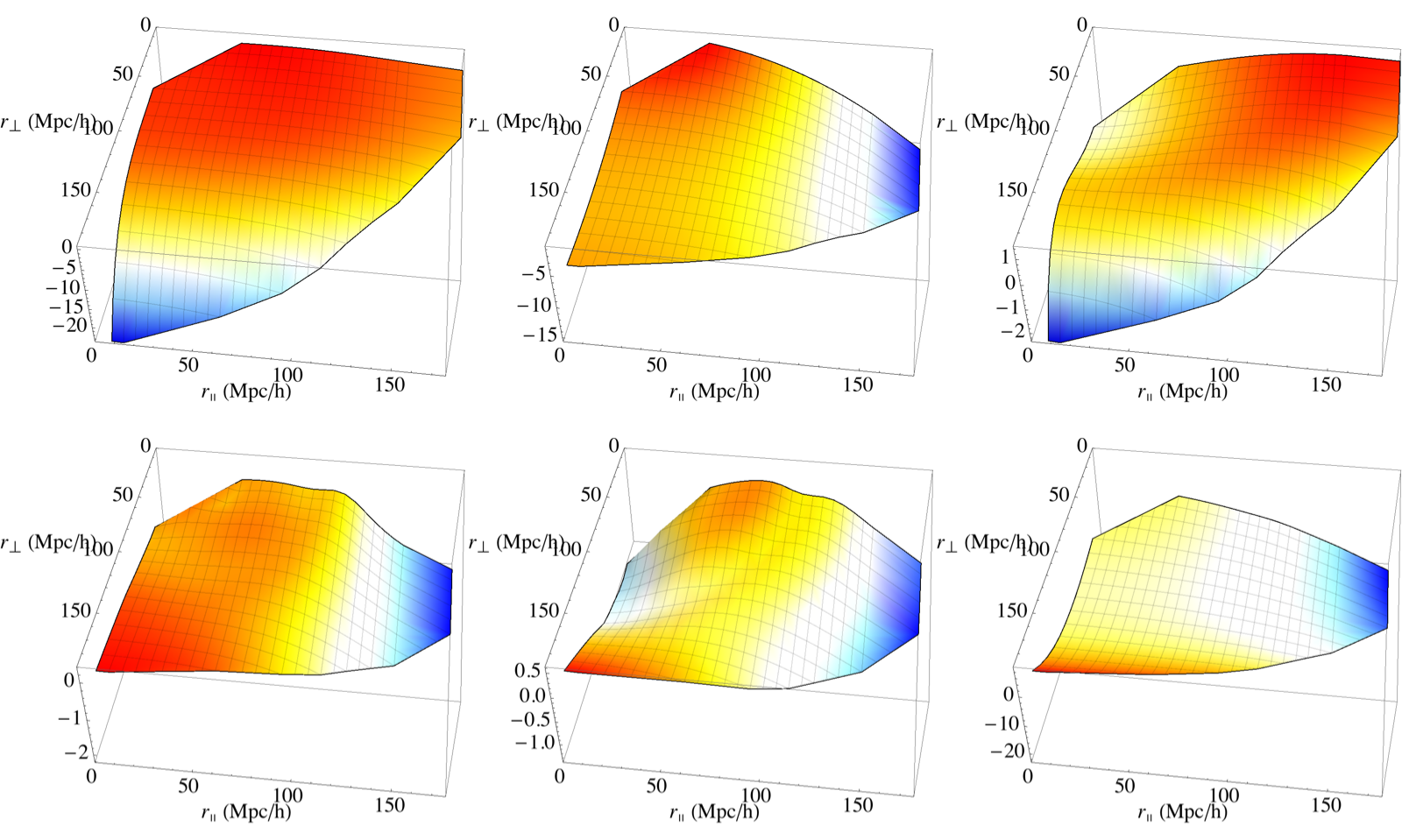}
\caption{Best fits of the six distortion models BB1--6 described in Table~\ref{table:bband-models} to DR9. Plots show the $r^2$-weighted distortion $\xi(r,\mu,z) - \xi_{\text{cosmo}}(r,\mu,z)$ for $z = 2.5$. Models BB1--3 are on the top row of plots, left to right, and BB4--6 are on the bottom row. The faint negative impression of the BAO feature visible for BB4 and BB5 reflects the fact that they model purely multiplicative distortions.}
\label{fig:bband-fits}
\end{center}
\end{figure}

We model broadband distortion using the following parametrization that combines the multiplicative ($B_m$) and additive ($B_a$) effects described above:
\begin{equation}
\xi(r,\mu,z) = \xi_{\text{cosmo}}(r,\mu,z)\cdot \left[ 1 + B_m(r,\mu,z)\right] +
B_a(r,\mu,z)\cdot \left( \frac{1+z}{1+z_0}\right)^{\gamma_{b^2}}
\end{equation}
with ($B_x$ represents either $B_m$ or $B_a$)
\begin{equation}
B_x(r,\mu,z) = \sum_{i=i_{\text{min}}}^{i_{\text{max}}} \sum_{j=j_{\text{min}}}^{j_{\text{max}}} \sum_{n=n_{\text{min}}}^{n_{\text{max}}} b^{(x)}_{i,j,n}\cdot
\left( \frac{r}{r_0} - \theta_i\right)^i \cdot L_{j}(\mu) \cdot \left( \frac{1+z}{1+z_0}\right)^n \; ,
\label{eqn:distortion-defn}
\end{equation}
where, nominally, $r_0 = 100$ Mpc/h and $z_0 = 2.25$, and
\begin{equation}
\theta_i =
\begin{cases}
0 & i \le 0\\
1 & i > 0\\
\end{cases}
\; .
\end{equation}
The integers $i_{\text{min}} \le i_{\text{max}}$, $0 \le j_{\text{min}} \le j_{\text{max}}$, and $n_{\text{min}} \le n_{\text{max}}$ determine the number of free parameters, and we normally restrict $j$ to even values. The $\ell = 0,2,4$ multipoles of the distorted correlation function are given by (arguments of $(r,z)$ are suppressed for clarity):
\begin{align}
\xi_0 =&\; \xi_{0,\text{cosmo}}\cdot\left( 1 + B_{0,m} \right) + \xi_{2,\text{cosmo}}\cdot\frac{1}{5}B_{2,m} +
\xi_{4,\text{cosmo}}\cdot\frac{1}{9}B_{4,m} + B_{0,a} \\
\xi_2 =&\; \xi_{0,\text{cosmo}}\cdot\left( 1 + B_{2,m} \right) + \xi_{2,\text{cosmo}}\cdot
\left( B_{0,m} + \frac{2}{7}B_{2,m} + \frac{2}{7}B_{4,m} \right) + \notag\\
&\quad\xi_{4,\text{cosmo}}\cdot\left( \frac{2}{7} B_{2,m} + \frac{100}{693} B_{4,m} + \frac{25}{143}B_{6,m} \right) + B_{2,a} \\
\xi_4 =&\; \xi_{0,\text{cosmo}}\cdot\left( 1 + B_{4,m} \right) + \xi_{2,\text{cosmo}}\cdot
\left( \frac{18}{35} B_{2,m} + \frac{20}{77}B_{4,m} + \frac{45}{143}B_{6,m} \right) + \notag\\
&\quad\xi_{4,\text{cosmo}}\cdot\left( B_{0,m} + \frac{20}{77} B_{2,m} + \frac{162}{1001} B_{4,m}
+ \frac{20}{143}B_{6,m} + \frac{490}{2431} B_{8,m} \right) + B_{4,a}
\end{align}
with distortion multipoles given by
\begin{equation}
B_{\ell,x}(r,z) =
\sum_{i=i_{\text{min}}}^{i_{\text{max}}} \sum_{n=n_{\text{min}}}^{n_{\text{max}}} b^{(x)}_{i,\ell,n}\cdot
\left( \frac{r}{r_0} - \theta_i\right)^i \cdot \left( \frac{1+z}{1+z_0} - 1 \right)^n \; .
\end{equation}
For the purposes of fitting DR9, we chose six different parameter configurations for broadband distortion, as summarized in Table~\ref{table:bband-models} and plotted in Figure~\ref{fig:bband-fits}.

\begin{table}[htb]
\begin{center}
\begin{tabular}{|l|rrr|rrr|}
\hline
& \multicolumn{3}{c|}{$B_m$} & \multicolumn{3}{c|}{$B_a$} \\
Name & $i$ & $j$ & $n$ & $i$ & $j$ & $n$ \\
\hline
BB1 & - & - & - & 0,1,2 & 0,2,4 & 0 \\
BB2 & - & - & - & -2,-1,0 & 0,2,4 & 0 \\
BB3 & - & - & - & 0,1,2 & 0,2 & 0,1 \\
BB4 & 0,1,2 & 0,2,4 & 0 & - & - & - \\
BB5 & 0,1,2 & 0,2 & 0,1 & - & - & - \\
BB6 & 0,1 & 0,2,4 & 0 & 0,1 & 0,2,4 & 0 \\
\hline
\end{tabular}
\end{center}
\caption{Broadband distortion models used to fit DR9. Models are labeled BB1-6 and columns show the range of indices in eqn.~(\ref{eqn:distortion-defn}) used in the multiplicative ($B_m$) and additive ($B_a$) components. Dashes indicate that a component is not used.}
\label{table:bband-models}
\end{table}

\subsection{Interpolated Models}
\label{sec:interpolated}

For the purposes of visualization and studying non-cosmological signatures in data, it is useful to fit using simple interpolated models with minimal assumptions. We describe here two such models: the first treats the correlation multipoles as (possibly independent) arbitrary interpolations through fixed arbitrary values of the separation, and the second treats corrections to a smoothed power spectrum as (possibly independent) arbitrary interpolations through equally spaced wave numbers covering a limited band.

For our first model, we pick $n$ fixed values of the separation $\{r_j\}$ and model the $r^2$-weighted correlation multipoles as:
\begin{equation}
r^2 \xi_{\ell}(r,z) = b^2_{\ell}(z)\,I_{m,n}(r,z_0;\{ r_j \})
\label{eqn:xi-model}
\end{equation}
where $I_{m,n}$ is an $m$-th order interpolation through the $n$ points $\{r_j\}$ evaluated at separation $r$ and referenced to the redshift $z_0$. Comparing with eqn.~(\ref{eqn:Ptilde}), we write
\begin{equation}
b^2_{\ell}(z) = b^2(z)\, C_{\ell}(\beta(z)) \; ,
\label{eqn:bsqell-defn}
\end{equation}
so that
\begin{equation}
I_{m,n}(r,z_0) = r^2\,\tilde{\xi}_{\ell,\text{cosmo}}(r,z_0)
\end{equation}
when $\xi_{\ell}(r,\mu,z)$ is purely cosmological.

\begin{figure}[htb]
\begin{center}
\includegraphics[width=6in]{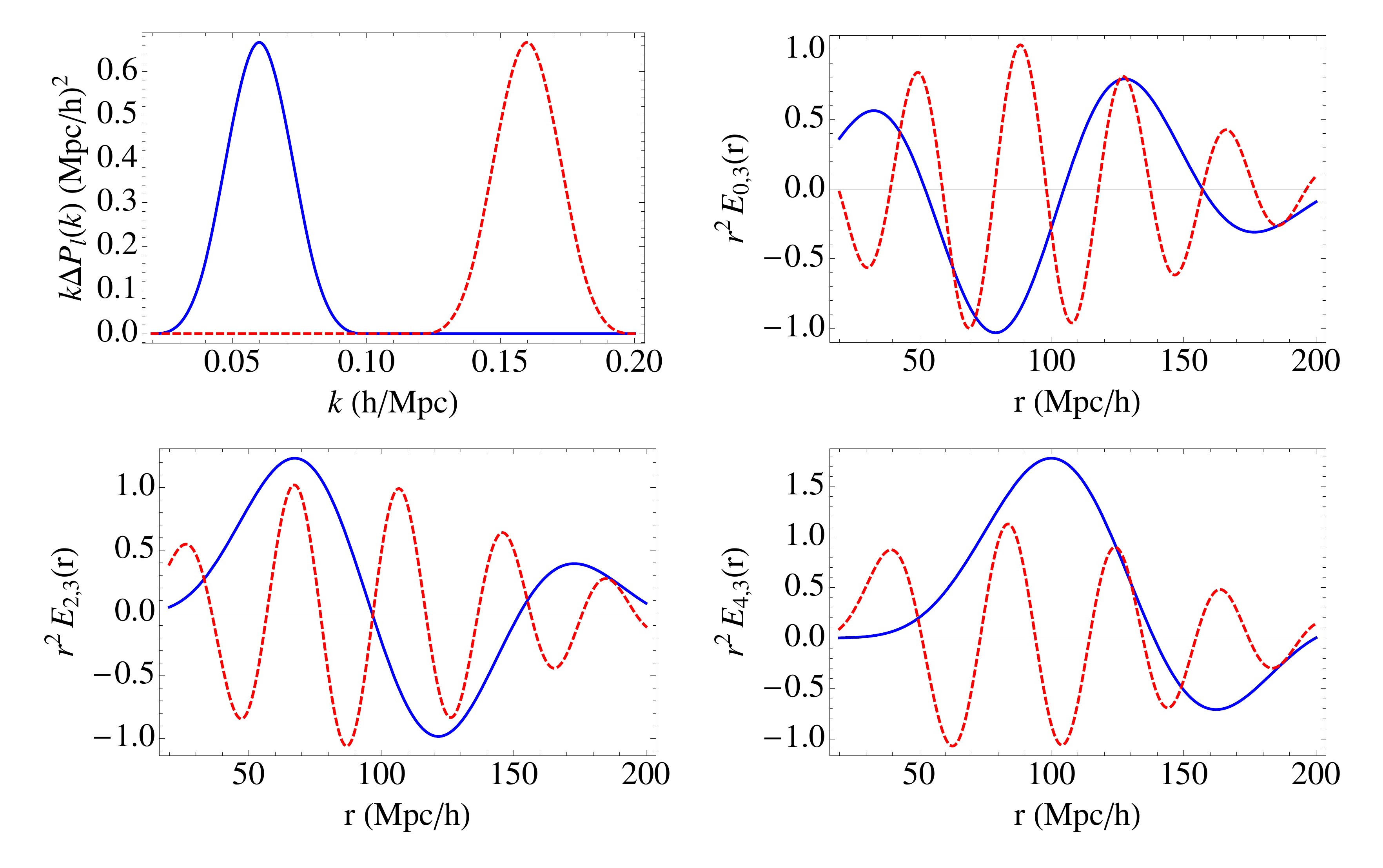}
\caption{Cubic ($m = 3$) basis spline contributions to the ($k$-weighted) power multipoles $k \Delta P_{\ell}(k)$ (top left) with their corresponding ($r^2$-weighted) contributions $E_{\ell,m}(r;k_j,\Delta k)$ with $\ell = 0, 2, 4$ (clockwise from top right) to the correlation function multipoles $\Delta\xi_{\ell}(r)$ with $k_{\text{lo}},k_{\text{hi}} = (0.02,0.2)$ h/Mpc, $n = 10$, and $j = 0$ (solid blue curves) and $j = 5$ (dashed red curves).}
\label{fig:b-splines}
\end{center}
\end{figure}

For our second model, we interpolate over a fixed band of $k$-space and therefore require some assumption about the power outside this band in order to predict $r$-space correlations. We adopt the following form for the multipoles of the power spectrum:
\begin{equation}
P_{\ell}(k,z) = b^2_{\ell}(z)\left[ \tilde{P}_{\text{smooth}}(k,z_0) + \Delta P_{\ell}(k,z_0) \right]
\label{eqn:pk-model}
\end{equation}
where $\tilde{P}_{\text{smooth}}(k,z_0)$ is a smoothed fiducial power spectrum, without any BAO features, modulated by functions $\Delta P_{\ell}(k,z_0)$ that are zero outside some interval $(k_{\text{lo}}, k_{\text{hi}})$, both referenced to $z_0$. The correlation multipoles corresponding to eqn.~(\ref{eqn:pk-model}) are obtained via eqn.~(\ref{eqn:xil-defn}):
\begin{equation}
\xi_{\ell}(r,z) = b^2_{\ell}(z)\left[ \tilde{\xi}_{\ell,\text{smooth}}(r,z_0) + \Delta\xi_{\ell}(r,z_0) \right]
\end{equation}
with
\begin{equation}
\Delta\xi_{\ell}(r,z_0) = \frac{i^\ell}{2\pi^2}\,\int_{k_{\text{lo}}}^{k_{\text{hi}}}\, k^2 j_{\ell}(k r) \Delta P_{\ell}(k,z_0)\,dk \; .
\end{equation}
We parameterize the $k$-weighted modulation as a linear combination of $m$-th order B-spline functions $B_m$:
\begin{equation}
k\cdot \Delta P_{\ell}(k,z_0) = \sum_{j=0}^{n-m-2} \, b_{\ell,j} B_m\left(\frac{k - k_j}{(m+1)\Delta k}\right)
\end{equation}
with $n$ uniformly spaced knots $k_j$ at:
\begin{equation}
k_j = k_{\text{lo}} + j\cdot \Delta k \quad , \quad \Delta k = \frac{k_{\text{hi}} - k_{\text{lo}}}{n - 1} \; .
\end{equation}
Using the property that $B_m(t)$ is only non-zero for $0 \le t \le 1$, we can write
\begin{equation}
\Delta\xi_{\ell}(r,z_0) = \frac{i^\ell}{2\pi^2}\,\sum_{j=0}^{n-m-2}\, b_{\ell,j}\, E_{\ell,m}(r; k_j, \Delta k)
\end{equation}
where
\begin{equation}
E_{\ell,m}(r; k_j, \Delta k) \equiv 
(m+1)\Delta k
\int_0^1\, k(t)\, j_{\ell}\left(k(t) r\right) B_m(t) dt
\end{equation}
with
\begin{equation}
k(t) = k_j + (m+1)\Delta k \cdot t
\end{equation}
can be solved analytically in terms of trigonometric functions and the sine integral. Figure~\ref{fig:b-splines} shows examples of $E_{\ell,m}$ for cubic ($m = 3$) B splines.

\section{Fitting Method}
\label{sec:fitting}

\begin{figure}[htb]
\begin{center}
\includegraphics[width=6in]{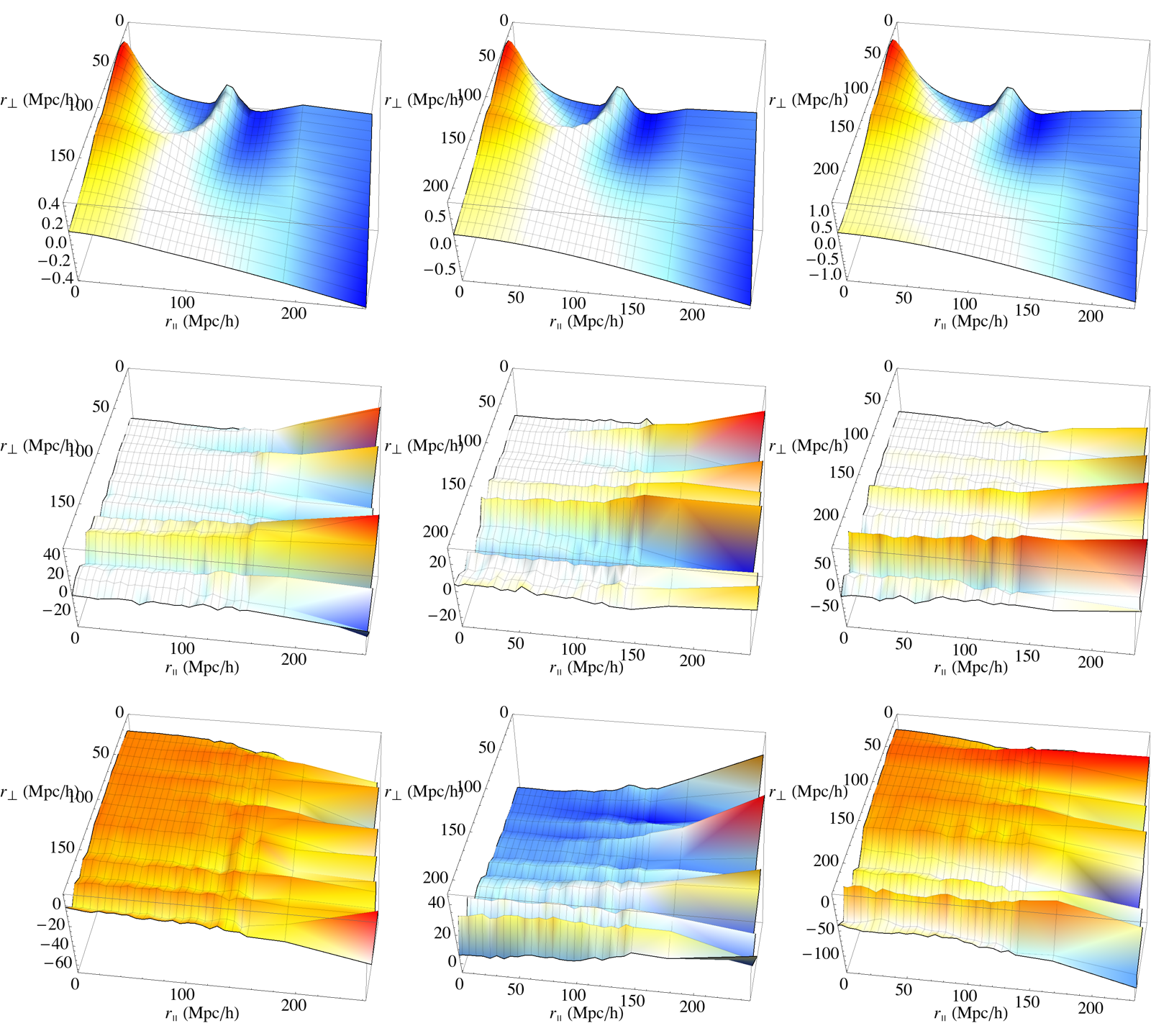}
\caption{Slices in $(r,\mu)$ of the estimated $r^2$-weighted correlation function for the fiducial cosmology (top row), simulated mocks (middle row), and BOSS data (bottom row). Grid lines show the actual sampling of $\xi_{ijk}$ used. From left to right, columns show redshifts of 2, 2.5, and 3. Physical coordinates are converted to comoving coordinates using the fiducial cosmology.}
\label{fig:rmu-slices}
\end{center}
\end{figure}

We fit the parameterized models of $\xi(r,\mu,z)$ described above to estimates $\xi_{ijk} \equiv \xi(r_{ijk},\mu_{ijk},z_k)$ of the correlation function specified on a 3D grid $\Delta v_i \otimes \Delta\theta_j \otimes z_k$ of physical coordinates (see Section~\ref{sec:physical-coords}), with
\begin{align}
r_{ijk} &= \sqrt{r_{\parallel}^2(\Delta v_i,z_k) + r_{\perp}^2(\Delta\theta_j,z_k)} \\
\mu_{ijk} &= r_{\parallel}(\Delta v_i,z_k) / r_{ijk} \; .
\end{align}
Our nominal coordinate grid consists of 28 unequally spaced points in $\Delta v$ covering $0 \le \Delta v < 0.083 c$ (see Figs.~\ref{fig:coords} and \ref{fig:rmu-slices}), 18 equally spaced points in $\Delta\theta$ spanning $5$--$175$ arcminutes, and three redshift values (2, 2.5, 3), for a total of $N = 1512$ grid points. Each set of correlation-function estimates $\xi_{ijk}$ is accompanied by estimated covariances \begin{equation}
\langle \xi_{ijk}\xi_{i'j'k'}\rangle - \langle\xi_{ijk}\rangle \langle\xi_{i'j'k'}\rangle
\end{equation}
for each physical coordinate pair $(ijk)$ and $(i'j'k')$. In the following, we use the notation $\mathbf{d}$ to refer to a vector of $N$ correlation function estimates $\xi_{ijk}$, and write $C$ for the corresponding $N\times N$ covariance matrix. See ref.~\cite{Slosar2013} for details on the estimates $\mathbf{d}$ and $C$ obtained from BOSS DR9 and accompanying simulated mock data that we use below. Appendix~\ref{sec:public} provides instructions for downloading these estimates as well as the software necessary to reproduce the main results provided here and in ref.~\cite{Slosar2013}. We define a standard chi-square in terms of a parameter vector $\boldsymbol{\theta}$ and theory prediction $\boldsymbol{\mu}(\boldsymbol{\theta})$ as
\begin{equation}
\chi^2(\boldsymbol{\theta}) \equiv \left(\boldsymbol{d} - \boldsymbol{\mu}(\boldsymbol{\theta})\right)^t
C^{-1} \left(\boldsymbol{d} - \boldsymbol{\mu}(\boldsymbol{\theta})\right) \; .
\label{eqn:chisq-defn}
\end{equation}

\subsection{SDSS-III Data Release 9 Inputs}

\begin{figure}[htb]
\begin{center}
\includegraphics[width=6in]{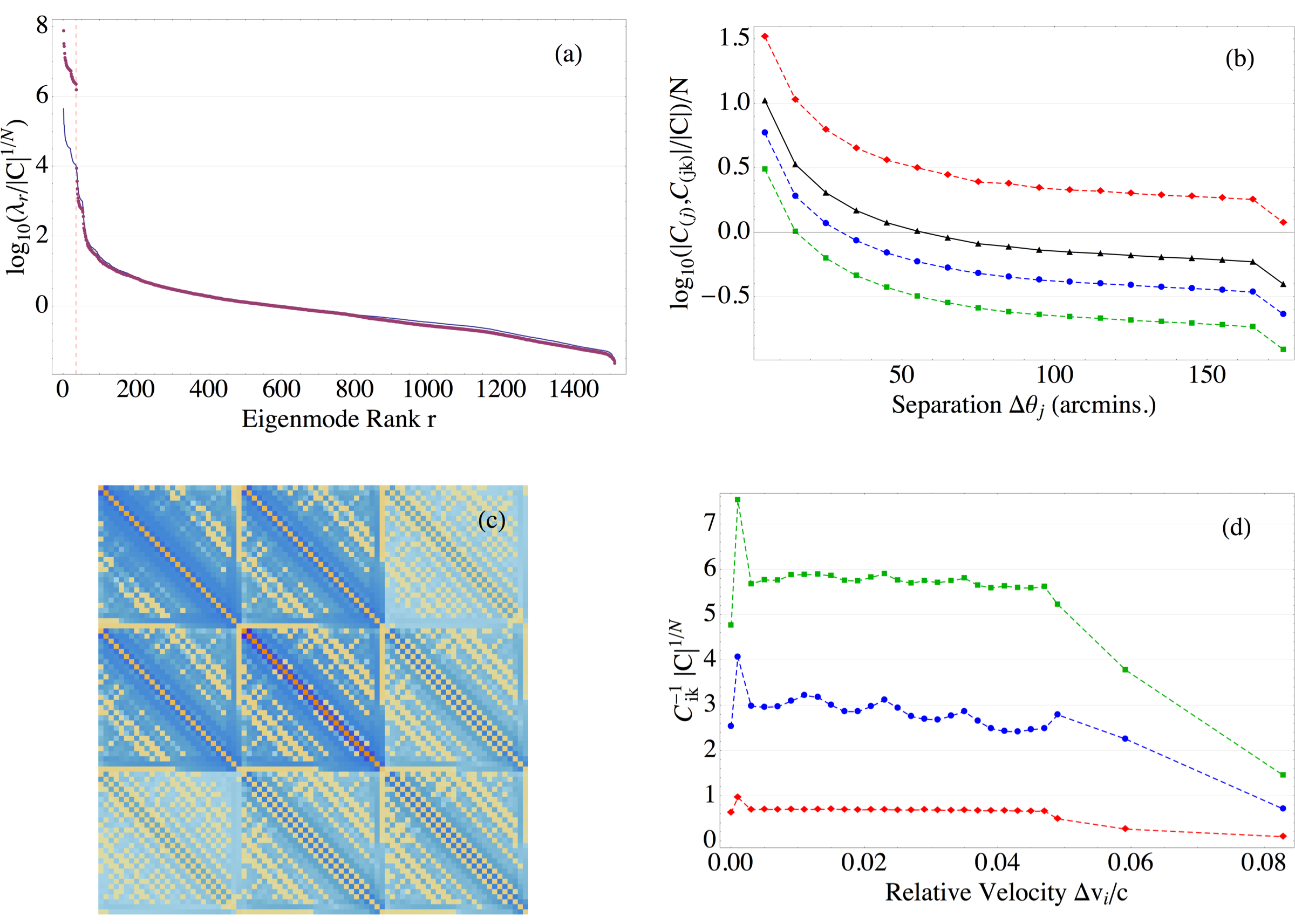}
\caption{Structure of the estimated covariance matrix $C$. The top-left panel (a) shows the ordered eigenmodes $\lambda_r$ of $C$ in data (blue curve) and simulation (points). The top-right panel (b) shows the relative normalization of the 18 block-diagonal submatrices $C_{(j)}$ as a function of separation $\Delta\theta_j$ (black), as well as the relative normalization of the three $28\times 28$ sub-matrices $C_{(jk)}$ corresponding to redshifts 2 (blue circles), 2.5 (green squares), and 3 (red diamonds), averaged over relative velocities $\Delta v_i$. The bottom-left panel (c) shows the common structure of the 18 $C^{-1}_{j}$ blocks for each separation $\Delta\theta_j$. The bottom-right panel (d) shows the relative inverse covariance along the diagonal of this common block, as a function of relative velocity $\Delta v_i/c$, for redshifts 2 (blue circles), 2.5 (green squares), and 3 (red diamonds).}
\label{fig:cov-structure}
\end{center}
\end{figure}

Figure~\ref{fig:rmu-slices} compares the data vectors for theory, simulated BOSS DR9 data, and actual BOSS DR9 data. The signal is visually obscured in the data by large-amplitude modes at fixed separation angles $\Delta\theta \sim r_{\perp}$. However, these modes have negligible impact on our ability to measure BAO parameters since they have very small weights (by construction) in the covariance matrix (see Figure~\ref{fig:cov-structure}). The value of $\log_{10}|C|/N$ provides a rough measure of the overall (i.e., not BAO specific) signal to noise ratio of a sample. We find values of -8.5 and -8.3 for data and mocks, respectively, with $N = 1512$, which indicates that the errors in our simulated data are about 30\% larger than in real data. The signal to noise ratio is largest at $z = 2.5$, with errors about 40\% larger at $z = 2$ and three times larger at $z = 3$.

Figure~\ref{fig:cov-structure} shows the structure of our covariance estimates, which is essentially the same for simulation and data after accounting for the $\sim 30$\% normalization difference. The first 36 eigenmodes have artificially large eigenvalues (see Figure~\ref{fig:cov-structure}a) due to the template marginalization procedure described in ref.~\cite{Slosar2013}, which accounts for the largest-amplitude modes visible in Figure~\ref{fig:rmu-slices}. We expect correlation function estimates to be uncorrelated between different separation angles and therefore impose this constraint on our estimated inverse covariance matrix reducing both $C$ and $C^{-1}$ to a block diagonal form consisting of 18 sub-matrices, $C_{(j)}$ and $C^{-1}_{(j)}$, with dimensions $84\times 84$. Each of these submatrices has the same structure, shown in Figure~\ref{fig:cov-structure}c for $C^{-1}_{(j)}$, but a different normalization, $|C_{(j)}|$ or $|C^{-1}_{(j)}| = |C_{(j)}|^{-1}$, depending on its separation $\Delta\theta_j$ (solid curve in Figure~\ref{fig:cov-structure}b). Each of the three $28\times 28$ sub-matrices on the diagonal, corresponding to a fixed separation $\Delta\theta_j$ and redshift $z_k$, also have the same structure with normalizations shown as the dashed curves in Figure~\ref{fig:cov-structure}b. Values of the inverse covariance along the diagonal are roughly independent of $\Delta v_i$ and only depend on $\Delta\theta_j$ via the factor $|C_{(j)}|$, as plotted in Figure~\ref{fig:cov-structure}d where the decreasing values at large $\Delta v_i$ correspond to increasing errors due to limited statistics.

\begin{figure}[htb]
\begin{center}
\includegraphics[width=6in]{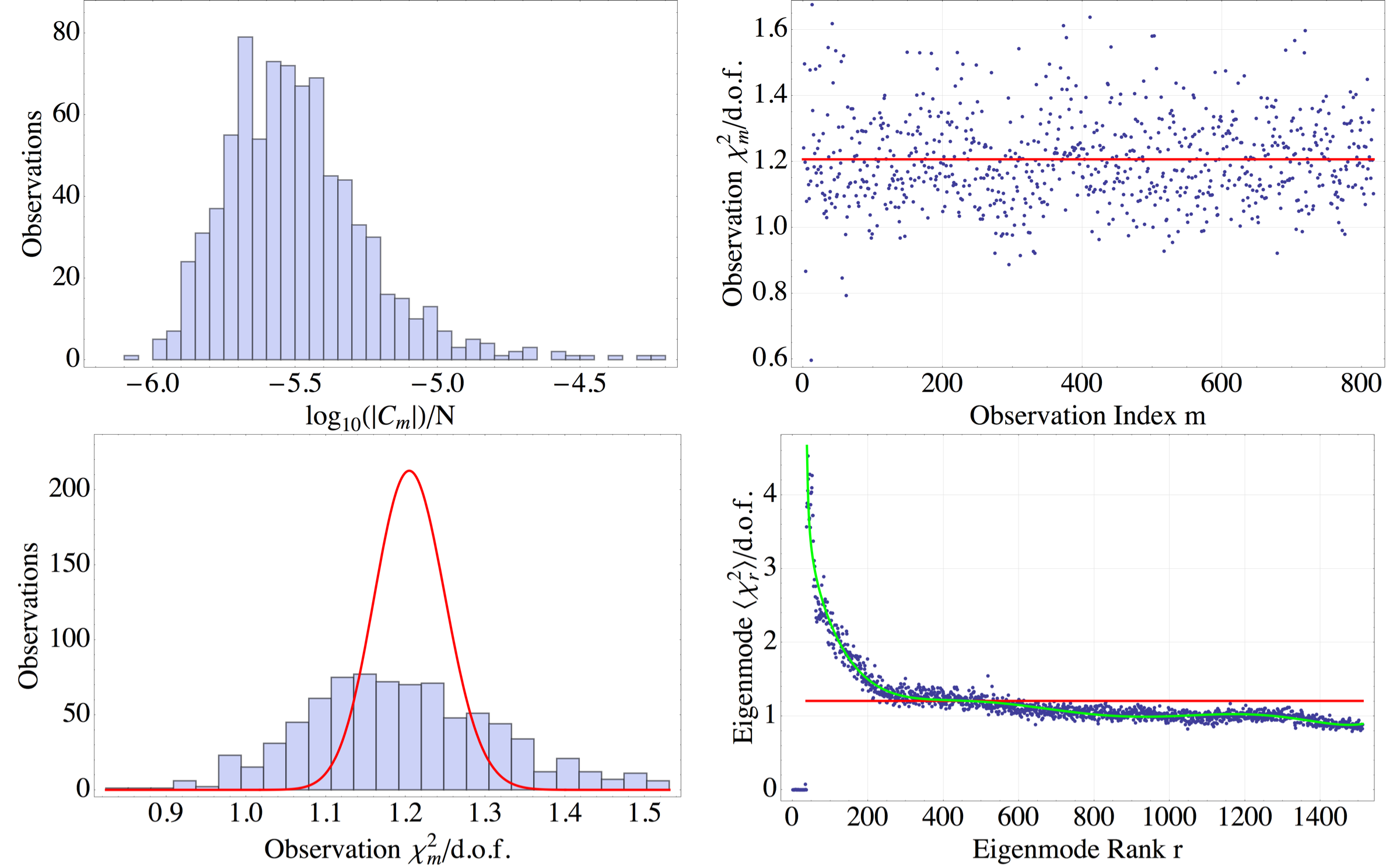}
\caption{Covariance studies of the 817 independent correlation function estimates in BOSS DR9. The top-left panel shows the distribution of covariance weights $\log_{10}(|C_j|)/N$ for each observation $j$. The $\chi^2_j$ values for each of the 817 observations are shown in the top-right panel as a function of observation index $j$, which is roughly ordered by date of observation, compared with their mean of 1.2. The bottom-left panel shows the distribution of normalized $\chi^2_j$, defined in eqn.~(\ref{eqn:chi2j-defn}), for 1512 degrees of freedom, compared with their expected distribution (red curve) for Gaussian statistics with an overall normalization factor of 1.2 applied. The bottom-right panel shows the evolution of $\langle\chi^2_r\rangle$ with eigenvalue rank $r$. Figure~\ref{fig:rescale-results} repeats the bottom two panels after applying the rescaling procedure described in the text.}
\label{fig:scale-analysis}
\end{center}
\end{figure}

Our correlation function estimates are actually provided as a set of $M$ quasi-independent data vectors $\mathbf{d_m}$ and covariance matrices $C_m$, corresponding to angular partition of the BOSS survey footprint into individual observing plates~\cite{Slosar2013}. For fits to DR9, we have 
$M = 817$ plates. We perform a weighted combination of observations, assumed to be independent, to obtain the final $\mathbf{d}$ and $C$ used for fitting
\begin{equation}
C^{-1} = \sum_{m=1}^M\, C_{m}^{-1} \quad , \quad
\mathbf{d} = C \, \sum_{m=1}^M C_{m}^{-1}\, \mathbf{d}_m \; .
\label{eqn:combine-obs}
\end{equation}
The effective signal to noise ratio of individual observations spans about an order of magnitude (see Figure~\ref{fig:scale-analysis}) due to varying observing conditions, so it is important that observations are correctly weighted in the combination. However, an overall normalization error in the estimated covariances $C_m$ still leads to a correctly weighted combination, even if the individual $C_m$ have different structures.

\subsection{Internal Covariance Tests and Refinement}
\label{sec:covariance-tests}

We exploit the large number of observations available to perform some internal cross checks on the individual $C_m$ and obtain improved estimates, as described below. First, we compare each observation $\mathbf{d}_m$ to the combined observations using the chi-square statistic
\begin{equation}
\chi^2_m \equiv (\mathbf{d}_j - \mathbf{d})^t (C_m - C)^{-1} (\mathbf{d}_j - \mathbf{d})
\label{eqn:chi2j-defn}
\end{equation}
where $C_m - C$ is the expected covariance of $\mathbf{d}_m - \mathbf{d}$
\begin{align}
\langle (\mathbf{d}_m - \mathbf{d})(\mathbf{d}_m - \mathbf{d})^t \rangle
&= \langle\mathbf{d}_m\mathbf{d}_m^t\rangle - 2 \langle\mathbf{d}_m\mathbf{d}^t\rangle + \langle\mathbf{d}\mathbf{d}^t\rangle \notag\\
&= C_m - 2 C + \sum_{m'} C C_{m'}^{-1} C \notag\\
&= C_m - C \; ,
\end{align}
and we have assumed that each $\mathbf{d}_m$ is a Gaussian realization of some common true correlation function $\mathbf{d}(\boldsymbol{\theta}_0)$ with additional noise $\mathbf{e}_m$ sampled from $C_m$
\begin{equation}
\mathbf{d}_m = \mathbf{d}(\boldsymbol{\theta}_0) + \mathbf{e}_m \; .
\end{equation}
Figure~\ref{fig:scale-analysis} shows that, on average, our covariance estimates $C_m$ underestimate the actual variance seen in the data vectors $\mathbf{d}_m$ by about 20\%, but that this factor is independent of when the observation was taken. We then search for a correlation between the degree of underestimation and the amount of predicted variance by diagonalizing each $C_m - C$
\begin{equation}
C_m - C = X_m \Lambda_m X_m^t
\end{equation}
where $\Lambda_m$ is a diagonal matrix of the $N = 1512$ eigenvalues ranked in increasing size (larger predicted variance)
\begin{equation}
(\Lambda_m)_{rs} = \delta_{rs}\,\lambda_{m,r} \; ,
\end{equation}
and $X_m$ is a matrix whose columns are the corresponding eigenvectors of $C_m - C$. We can then rewrite $\chi^2_j$ as
\begin{equation}
\chi^2_m = \sum_{r} \chi^2_{m,r}
\end{equation}
where 
\begin{equation}
\chi^2_{m,r} \equiv \lambda_r^{-1} u_{m,r}
\end{equation}
and $u_{m,r}$ are the components of $\mathbf{d}_m - \mathbf{d}$ in the eigenvector basis for $C_m - C$
\begin{equation}
\mathbf{u}_m = X^t_m (\mathbf{d}_m - \mathbf{d}) \; .
\end{equation}
Each $\chi^2_{m,r}$ should be chi-square distributed with one degree of freedom (if our assumptions of Gaussianity are valid) so we use their mean over the $M = 817$ observations as a function of eigenvalue rank $r$
\begin{equation}
\langle\chi^2_r\rangle \equiv \frac{1}{M}\, \sum_r \chi^2_{j,r}
\end{equation}
to provide another internal test, where we expect $\langle\chi^2_r\rangle \simeq 1$ for an eigenmode whose variance $\lambda_r$ is correctly estimated. The results show (see Figure~\ref{fig:scale-analysis}) that most of the 20\% underestimation is due to the first $\sim250$ eigenmodes with the largest predicted variances (but not including the 36 modes with artificially large variance mentioned above, which appear here with very small values of $\langle\chi^2_r\rangle$).

\begin{figure}[htb]
\begin{center}
\includegraphics[width=6in]{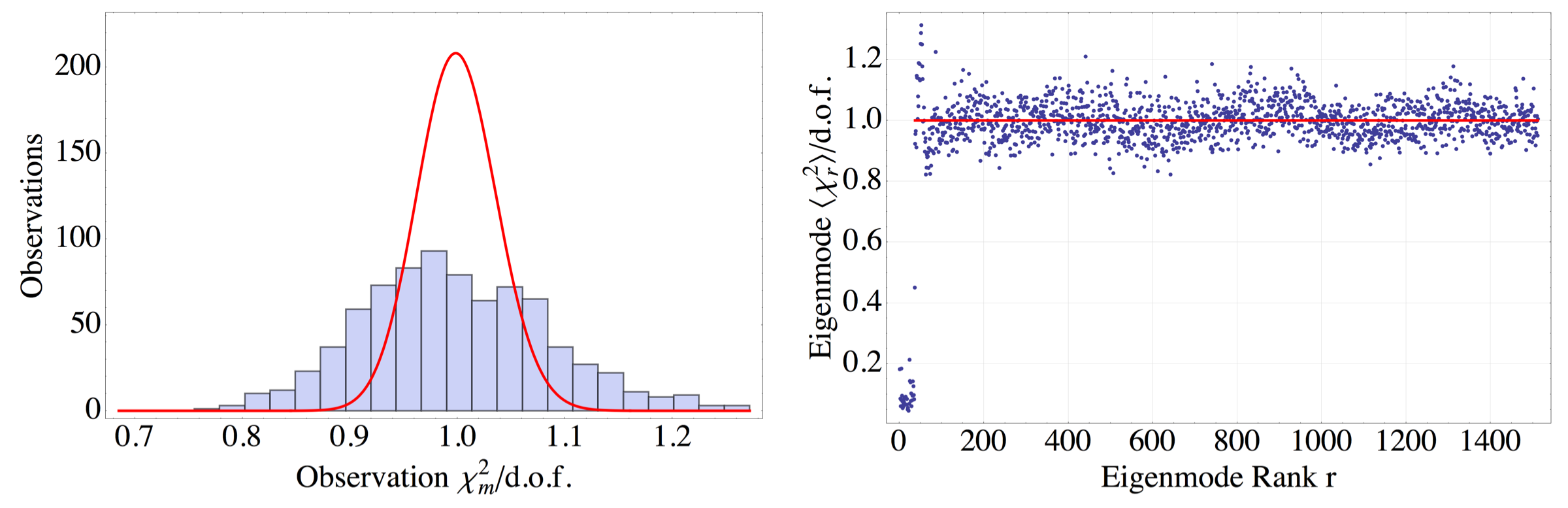}
\caption{Results of the rescaling procedure described in the text applied to BOSS DR9. Plots shows the same quantities as the bottom row of Figure~\ref{fig:scale-analysis} after applying the rescaling procedure described in the text.}
\label{fig:rescale-results}
\end{center}
\end{figure}

We can partially correct for the deviations with eigenvalue rank shown in Figure~\ref{fig:scale-analysis} by independently rescaling each eigenmode $r$ of $C_m$ by $\langle\chi^2_r\rangle^{-1}$. This procedure is not exact, however, since $\langle\chi^2_r\rangle^{-1}$ is derived from $C_m - C$ rather than $C_m$ (but we always have $|C_m| \gg |C|$) and because each $C_m$ generally has different eigenvectors (but we expect similar eigenvectors since the $C_m$ differ primarily in their normalizations). The results of this rescaling procedure are shown in Figure~\ref{fig:rescale-results} and demonstrate that all eigenmodes now have internally consistent $\mathbf{d}_m$ fluctuations and covariance estimates $C_m$. Two caveats to this procedure are that we do not rescale the 36 previously marginalized modes and that we impose the expected block-diagonal structure on the rescaled $C_m$, eliminating any off-diagonal contributions introduced by numerical round-off errors. Note that the distribution of the $\chi^2_m$ statistic has improved somewhat, compared with the Gaussian expectation, after rescaling. Finally, we combine the rescaled $C_m$ using eqn.~(\ref{eqn:combine-obs}) to obtain the final $\mathbf{d}$ and $C$ used for the fits described below (and in Figs.~\ref{fig:rmu-slices} and \ref{fig:cov-structure}).

We use bootstrap sampling in two different ways in our analysis. First, we boostrap our 817 estimates of the data vector $\mathbf{d}_m$ to obtain an estimate of their combined covariance that is independent of the $C_m$, as a cross-check of our combination using eqn.~(\ref{eqn:combine-obs}). With $M = 817$ and $N = 1512$, This approach is only feasible because of the $18\times 84^2$ block-diagonal structure of $C$: we effectively bootstrap the $84(84+1)/2 = 3570$ elements of each block $C_{(j)}$ independently using 817 sub-vectors of length 84. We compare fit results obtained with the calculated combination $C$ and the corresponding bootstrap estimate of $C$ to check the validity of our calculated $C$.

The second type of bootstrap analysis we perform is to fit a large number of bootstrap samples and then compare the resulting distributions of best-fit parameter values with the parabolic errors calculated from the likelihood surface of the combined fit using $\mathbf{d}$ and $C$. To the extent that the errors in our fit parameters are not Gaussian, we expect some discrepancy and then prefer the bootstrap estimates. Since our observations span an order of magnitude in weights $|C_m|$, a sampling with replacement generally yields a set of observations whose combined covariance does not match the combined covariance of all observations. Therefore, some care is required in assigning the total covariance used to fit each bootstrap sample since, although we ignore the resulting fit errors, we still require a correct relative weighting over the physical coordinate grid in eqn.~(\ref{eqn:chisq-defn}).

We generalize the combination method of eqn.~(\ref{eqn:combine-obs}) for bootstrap sampling to
\begin{equation}
\tilde{C}^{-1} = \sum_{m=1}^{M}\, n_m C_{m}^{-1} \quad , \quad
\mathbf{d} = \tilde{C} \, \sum_{m=1}^{M} n_m C_{m}^{-1}\, \mathbf{d}_m \; ,
\label{eqn:combine-bootstrap}
\end{equation}
where $n_m \ge 0$ is the number of repetitions of observation $m$ and $\sum_{m=1}^{M} n_m = M'$ is the bootstrap sample size (usually $M = M'$). The resulting $\tilde{C}$ is not the covariance of $\mathbf{d}$ since it incorrectly reduces the variance for double-counted observations. Instead, the correct covariance $C$ to use is
\begin{equation}
C = \tilde{C} D^{-1} \tilde{C}
\end{equation}
with
\begin{equation}
D^{-1} \equiv \sum_{m=1}^{M}\, n_m^2 C_m^{-1} \; .
\end{equation}
Note that eqn.~(\ref{eqn:combine-bootstrap}) reduces to eqn.~(\ref{eqn:combine-obs}) when each $n^2_m = n_m$, so that each sample is either not used, $n_m = 0$ (if $M' < M$), or used exactly once, $n_m = 1$. In the limit that all $C_m$ are identical, we find
\begin{equation}
\frac{\langle\tilde{\chi}^2\rangle}{\langle\chi^2\rangle} = \frac{M'+M-1}{M} \; ,
\end{equation}
where $\tilde\chi^2$ is the chi-squared that would be obtained using $\tilde{C}^{-1}$ instead of $C^{-1}$ in eqn.~(\ref{eqn:chisq-defn}). In the usual case of $M'=M$, we find a ratio $2 - 1/M \simeq 2$.

Before performing fits, we generally apply some final cuts to select a subset of the $N$ three-dimensional grid points. Cuts can either be in physical or (using the fiducial cosmology) comoving coordinates. We implement cuts by first combining all observations using eqn.~(\ref{eqn:combine-obs}), without cuts, and then eliminating elements from the final data vector $\mathbf{d}$, as well as the corresponding rows and columns of $C$. Note that the resulting $C^{-1}$ is different from what we would have obtained by eliminating rows and columns from $C^{-1}$ directly. Our nominal final cuts on physical coordinates are $0.003 < \Delta v_i/c < 0.083$ and $5 \le \Delta\theta_j \le 165$ arcminutes. We also cut on the co-moving separation $50 < r < 190$ Mpc/h. After all final cuts, the size of our data vector is reduced from $N = 3\times 504 = 1512$ to 341 ($z = 2$) + 310 ($z = 2.5$) + 290 ($z=3$) = 941.

\section{Results}
\label{sec:results}

In this section, we present results related to the fitting methods and what they reveal about the BOSS DR9 Lyman-$\alpha$ forest sample. Cosmological fit results derived from the same sample are presented in the companion paper ref.~\cite{Slosar2013}.

\subsection{Parameter Sensitivities}

\begin{figure}[htb]
\begin{center}
\includegraphics[width=6in]{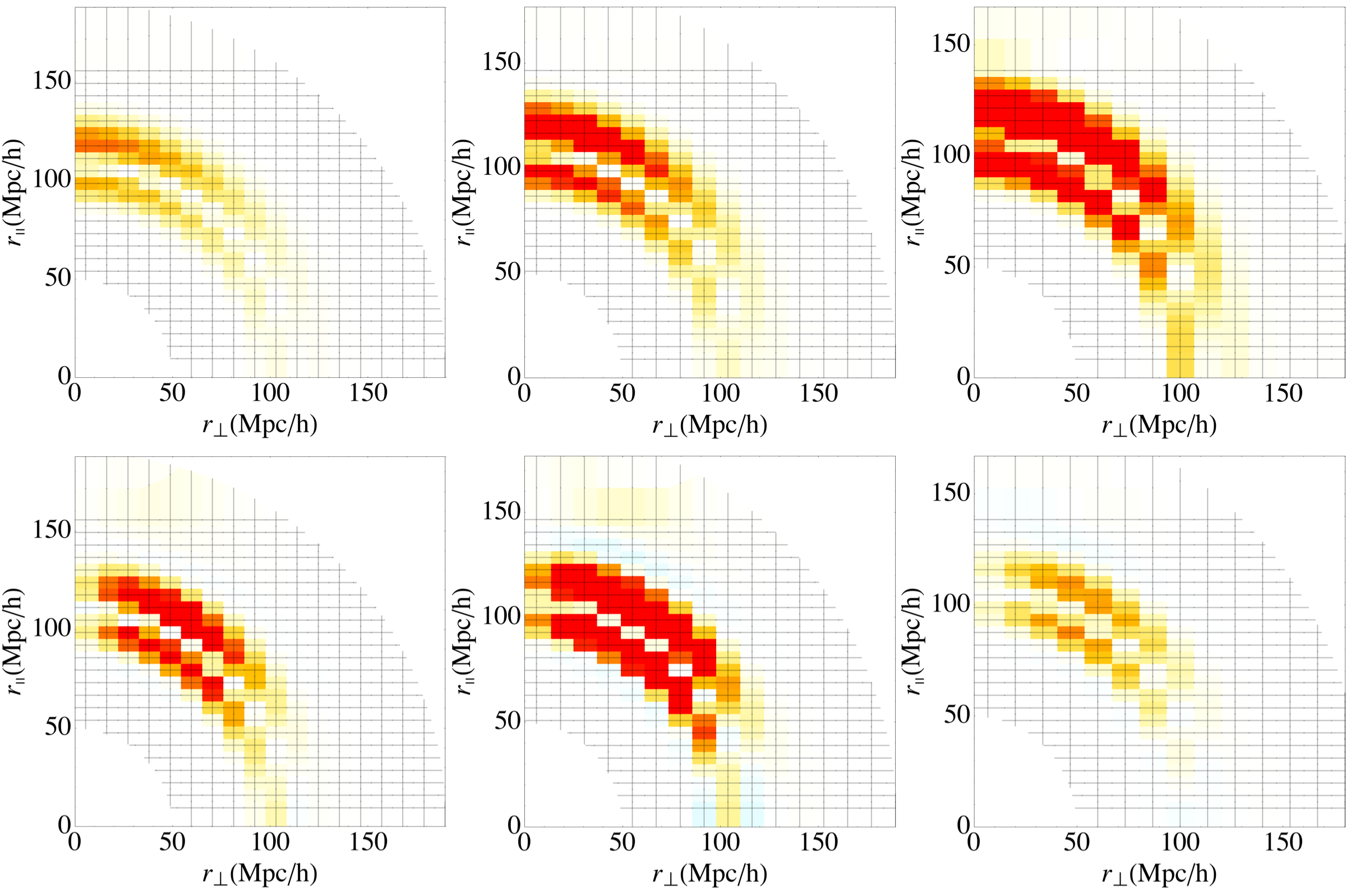}
\caption{Plots of the components of $\mathbf{F}_p$ for the isotropic scale parameter $\alpha_{\text{iso}}$. The top row shows the intrinsic model sensitivity using $C = |C|^{1/n}$, while the bottom row uses the combined covariance matrix for BOSS DR9. The three plots on each row show element values in the $(r_{\parallel},r_{\perp})$ for $z = 2$ (left), 2.5 (middle), and 3 (right), using the same color scale to preserve the relative magnitudes between redshifts.}
\label{fig:fisher-iso}
\end{center}
\end{figure}

\begin{figure}[htb]
\begin{center}
\includegraphics[width=6in]{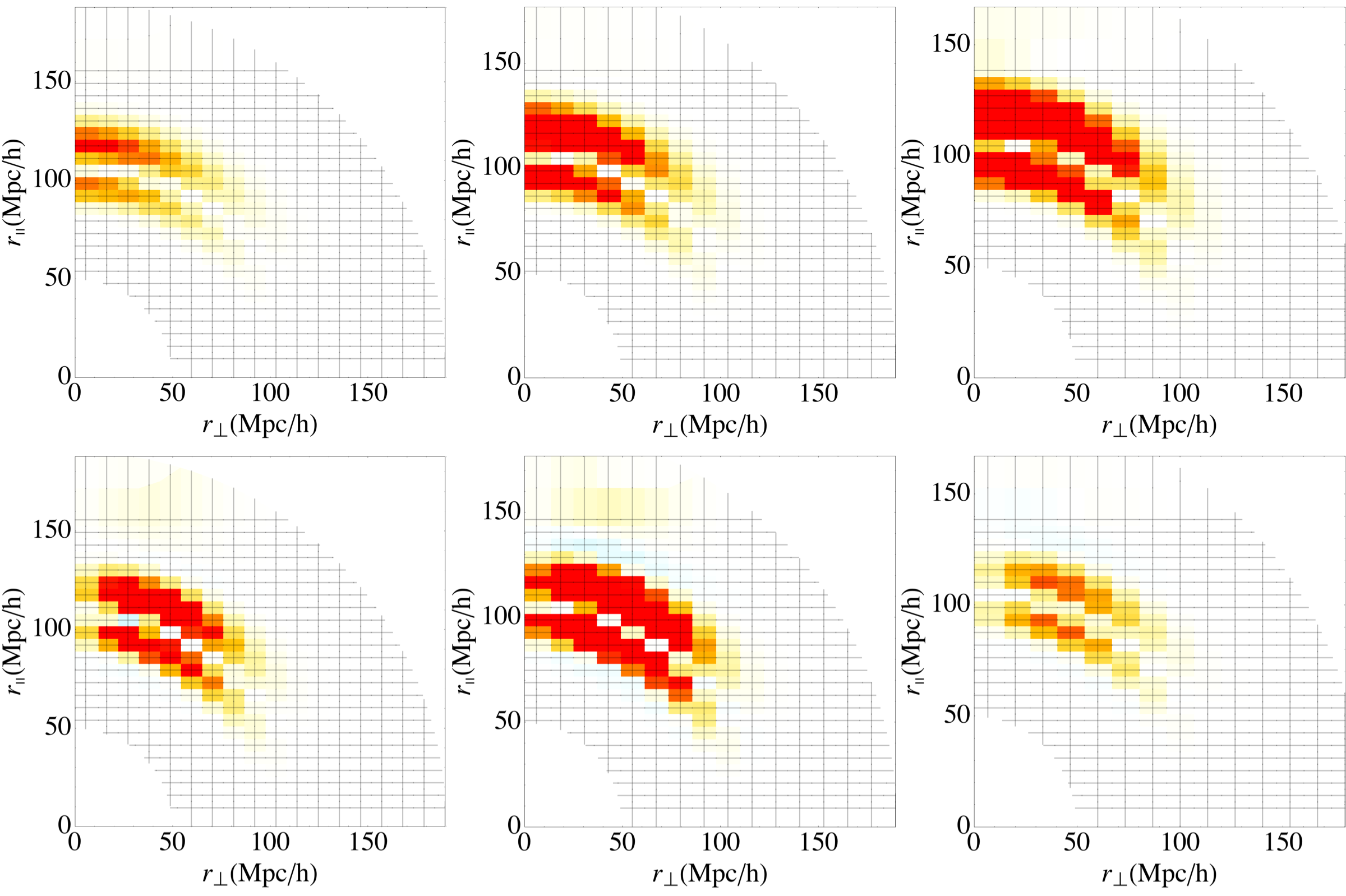}
\caption{Plots of the components of $\mathbf{F}_p$ for the anisotropic scale parameter $\alpha_{\parallel}$. See caption to Figure~\ref{fig:fisher-iso} for details.}
\label{fig:fisher-par}
\end{center}
\end{figure}

\begin{figure}[htb]
\begin{center}
\includegraphics[width=6in]{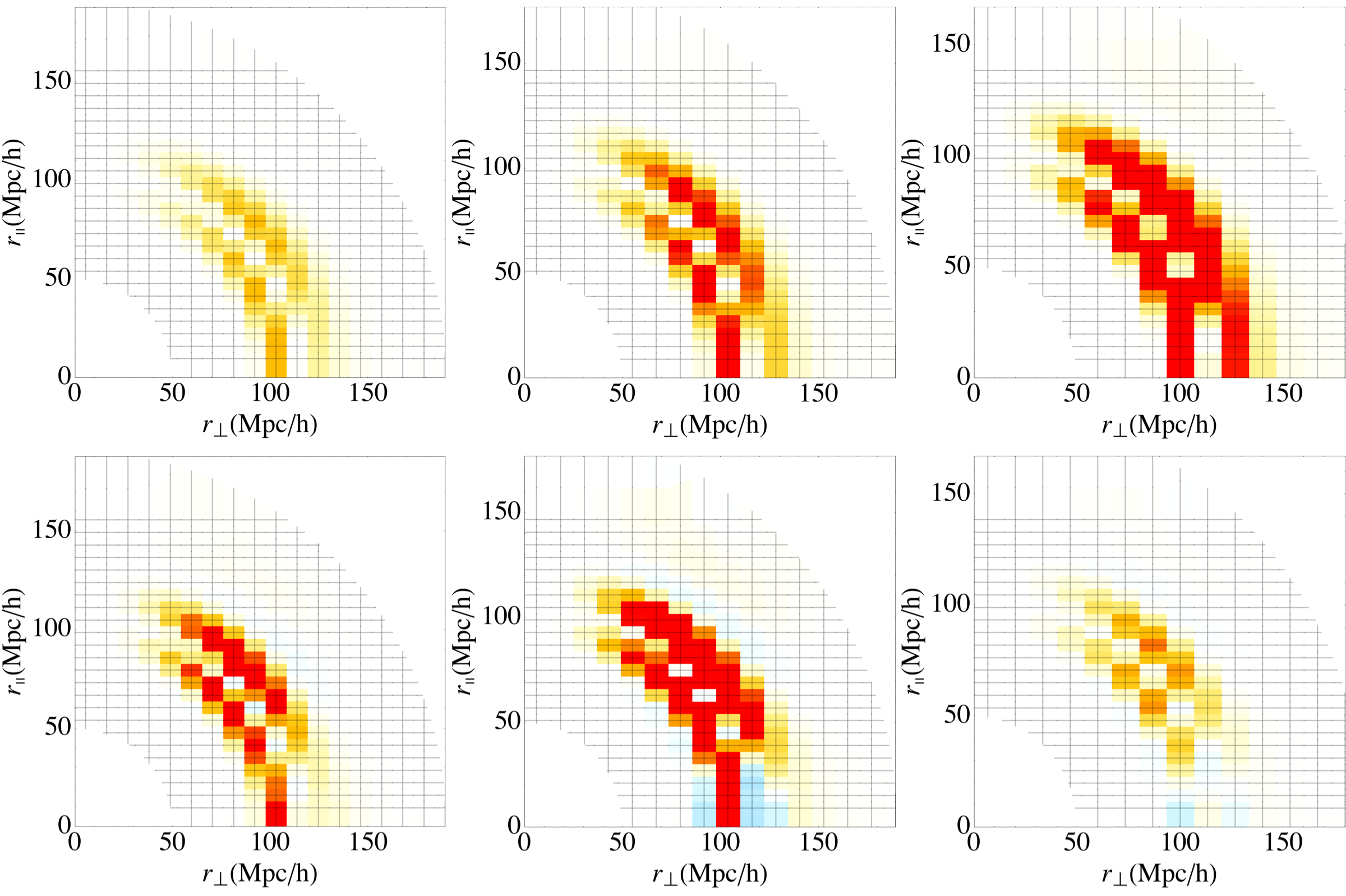}
\caption{Plots of the components of $\mathbf{F}_p$ for the anisotropic scale parameter $\alpha_{\perp}$. See caption to Figure~\ref{fig:fisher-iso} for details.}
\label{fig:fisher-perp}
\end{center}
\end{figure}

In order to quantify and visualize how a dataset constrains each parameter $p$, it is instructive to plot the vector
\begin{equation}
\mathbf{F}_p \equiv \mathbf{d}(\boldsymbol{\theta})_{,p} \circ \left( C^{-1} \mathbf{d}(\boldsymbol{\theta})_{,p} \right)
\end{equation}
where $\mathbf{d}(\boldsymbol{\theta})_{,p}$ is the partial derivative of the theory prediction at some point in parameter space $\boldsymbol{\theta}$, and $\circ$ represents the Hadamard (entrywise) product. The sum of the resulting components is the Fisher information for parameter $p$~\cite{1997ApJ...480...22T} that specifies the expected maximum-likelihood error $\sigma_p$ (in an ensemble-average sense) when all other parameter values are known
\begin{equation}
\sum_{ijk} F_{p,ijk} \simeq \sigma(p)^{-2} \; .
\end{equation}
The theory derivatives $\mathbf{d}(\boldsymbol{\theta})_{,p}$ encode the parameter sensitivity inherent to the model while the covariance $C$ encodes the performance of a particular analysis method. The quantity $\mathbf{F}_p$ is plotted for our baseline fit to BOSS DR9 for the isotropic scale parameter $\alpha_{\text{iso}}$ in Figure~\ref{fig:fisher-iso} and for the anisotropic scale parameters in Figures~\ref{fig:fisher-par} ($\alpha_{\parallel}$) and \ref{fig:fisher-perp} ($\alpha_{\perp}$).

The top row of Figure~\ref{fig:fisher-iso} uses $C = |C|^{1/n}$ to isolate the intrinsic model sensitivity, resulting in a Fisher error prediction of $\sigma(\alpha_{\text{iso}}) = 0.017$, with inverse-variance contributions of 12\%, 28\%, 60\% from $z =$ 2, 2.5, 3, respectively, due to the observed large increase in tracer bias $b^2(z)$ with redshift governed by the parameter $\gamma_{b^2}$. Information on the BAO scale comes predominantly from the forward direction due to the large observed value of the redshift-distortion parameter $\beta$, and is concentrated on two rings in co-moving separation on either side of the BAO peak, where the theory is most sensitive to the peak position. When the combined covariance $C$ is taken into account, the Fisher error prediction increases slightly to $\sigma(\alpha_{\text{iso}}) = 0.018$, and the region of peak sensitivity moves from $\mu = 1$ to $\mu \simeq 0.9$, mostly due to the trend shown in Figure~\ref{fig:cov-structure}(b). Intermediate redshifts now dominate the statistical power due to the distribution of forest pixel redshifts, with inverse-variance contributions of 29\%, 61\%, 10\% from $z =$ 2, 2.5, 3, respectively.

When floating the scales parallel and perpendicular to the line of sight, the Fisher predicted errors are $\sigma(\alpha_{\parallel}) =$ 0.024 (0.021) and $\sigma(\alpha_{\perp}) =$ 0.050 (0.047) for the combined analysis covariance $C$ (using $C = |C|^{1/n}$). Note that our actual covariance actually reduces the expected error on $\sigma(\alpha_{\perp})$, relative to $C = |C|^{1/n}$, because of the decreasing errors with separation angle shown in Figure~\ref{fig:cov-structure}b. The peak predicted sensitivities are $\mu \simeq 0.95$ for $\alpha_{\parallel}$ and $\mu \simeq 0.70$ for $\alpha_{\perp}$.

\begin{figure}[htb]
\begin{center}
\includegraphics[width=6in]{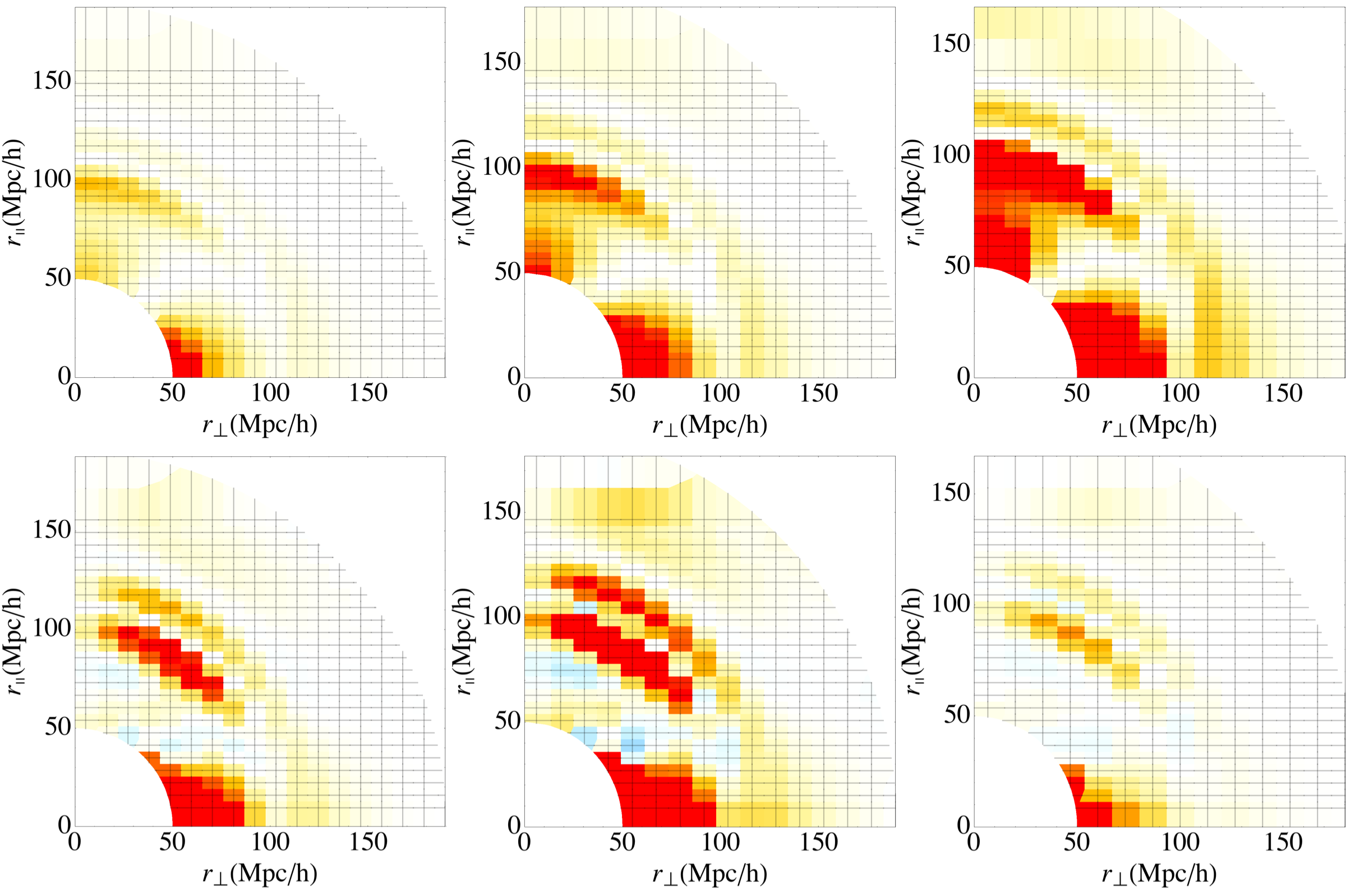}
\caption{Plots of the components of $\mathbf{F}_p$ for the isotropic scale parameter $\alpha_{\text{iso}}$, using $r'' = r'$ instead of $r'' = r$ in eqn.~(\ref{eqn:bband-defn}). See caption to Figure~\ref{fig:fisher-iso} for details.}
\label{fig:fisher-coupled}
\end{center}
\end{figure}

The results above were obtained with $r''=r$ in eqn.~(\ref{eqn:bband-defn}), so that scale factors are applied only to the peak position and not the cosmological broadband. If, instead, we use $r'' = r'$, the resulting Fisher components for an isotropic fit are shown in Figure~\ref{fig:fisher-coupled}, resulting in a smaller predicted error of $\sigma(\alpha_{\text{iso}}) = 0.010$ (0.013) for the combined analysis covariance $C$ (using $C = |C|^{1/n}$). Comparing with Figure~\ref{fig:fisher-iso}, we conclude that a fit using coupled transforms obtains a smaller error (when all other parameter values are fixed) by using information outside of the peak region. However, this benefit is offset by a larger dependence on the broadband distortion model, and the final errors are comparable when marginalizing over distortion parameters. For fits to DR9, we prefer to localize the BAO scale measurement to the peak region and minimize our dependence on the distortion model. With larger data samples and improved analyses, we expect to have better control of distortion effects and this choice should be revisited.

\subsection{Data Reductions}

The 1512 fit inputs $\xi_{ijk}$ already represent a substantial reduction of data from the millions of Lyman-$\alpha$ forest pixel pairs from which they are derived, while preserving essentially all of the cosmologically relevant information. We can take this process one step further using the interpolated models of Section~\ref{sec:interpolated} to reduce $\xi_{ijk}$ down to estimates of multipoles of the correlation function or power spectrum. There are two main challenges in this process: first, the input $\xi_{ijk}$ has large correlated errors that will propagate through and, second, the input $\xi_{ijk}$ is expected to include some broadband distortion that prevents direct comparison of reduced data with theory. Therefore, we first test our data reductions on a mock dataset based on our fiducial cosmology and with the same covariance as data, but no distortion, and then compare these to the corresponding results obtained from data.

We use three different types of model-independent fits for data reduction. The first fit estimates the $r^2$-weighted correlation function multipoles at $z = 2.4$ as a linear interpolation between 22 co-moving separations (5--140 in steps of 5, followed by 150, 160, and 180 Mpc/h) using eqn.~(\ref{eqn:xi-model}), for a total reduction from 1512 to 66 values. The second fit estimates the $k$-weighted power spectrum multipoles at $z = 2.4$ as a set 20 of equally spaced band powers covering $k = 0.03$--0.33 h/Mpc with $\Delta k = 0.015$ h/Mpc, using eqn.~(\ref{eqn:pk-model}) with 0-th order B-splines, for a total reduction from 1512 to 60 values. Our final model assumes that the $P(k)$ multipoles are identical except for the normalization factor eqn.~(\ref{eqn:bsqell-defn}), which is expected to be a good approximation if broadband distortion is either small or else experiences similar redshift-space distortion as the linear theory, and results in 20 final values. In all of our data reduction fits, we assume $\beta(z_0) = 1.4$, $\gamma_{b^2} = 3.8$, and $\gamma_{\beta} = 0$, and our results are normalized using $b(z_0) = -0.183$. We have made the results of these data reductions publicly available (see Appendix~\ref{sec:public}).

\begin{figure}[htb]
\begin{center}
\includegraphics[width=6in]{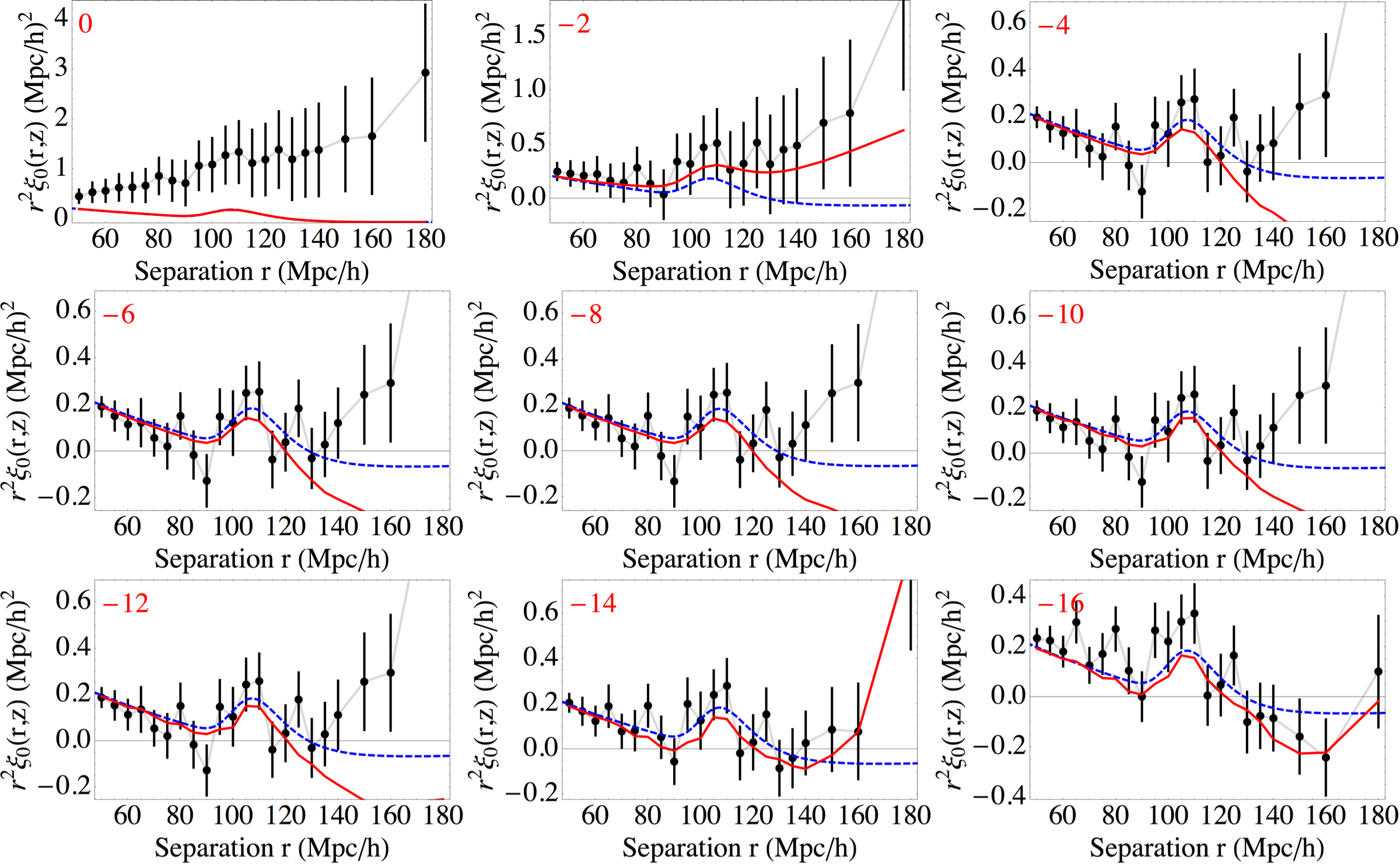}
\caption{Data reduction fit to an un-distorted mock with an increasing number of the largest eigenmodes (shown in the top-left corner of each plot) projected out of the resulting parameter covariance matrix. Points show the $r^2$-weighted monopole $\xi_0(r,z)$ at $z = 2.4$ of the reduced data with errors calculated from the corresponding diagonal parameter covariance matrix elements. Curves show the fiducial model at $z = 2.4$ with (red) and without (blue, dashed) the projection distortion. Neither curve is a fit to the data points.}
\label{fig:xi-mock-fit}
\end{center}
\end{figure}

\begin{figure}[htb]
\begin{center}
\includegraphics[width=6in]{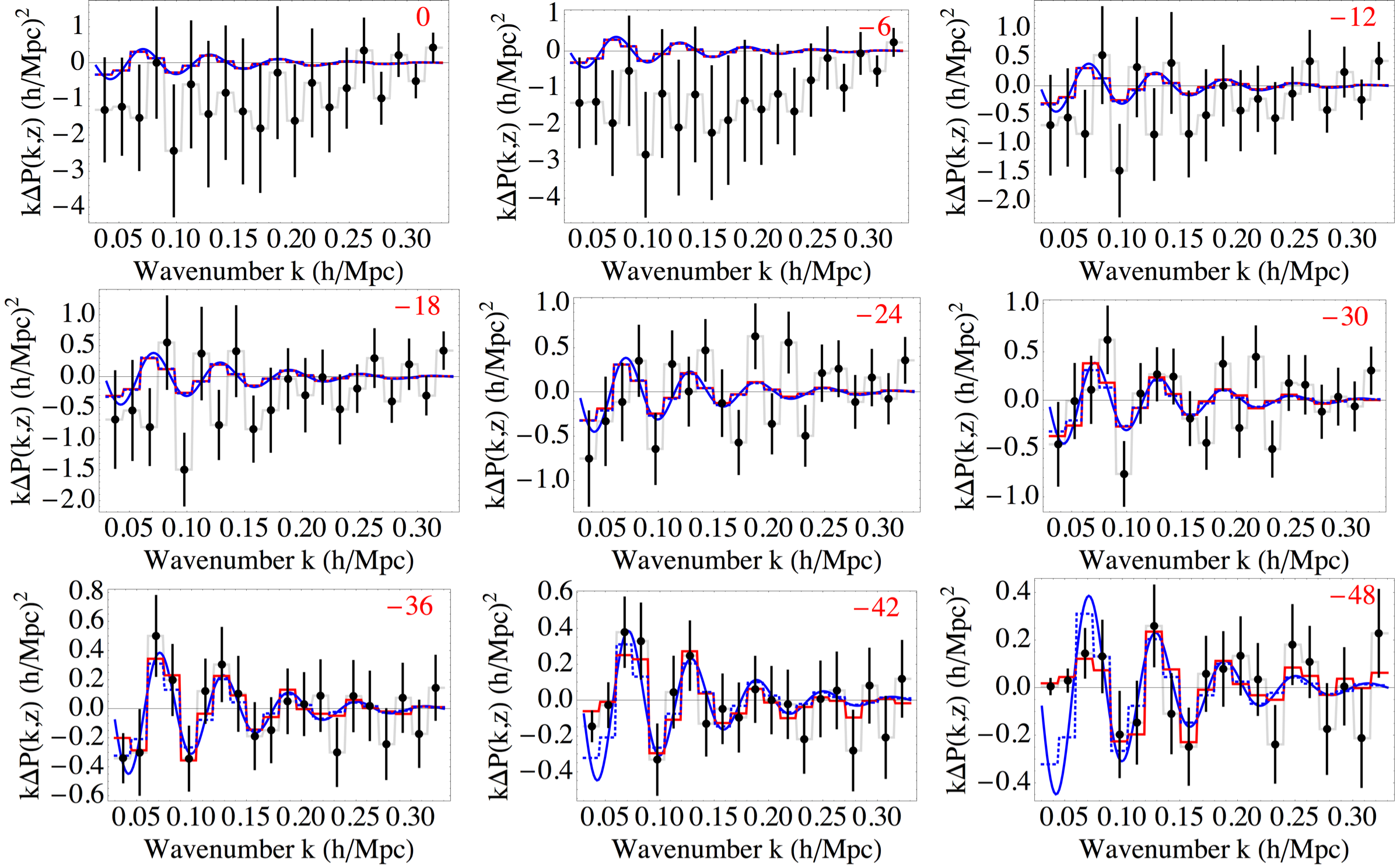}
\caption{Data reduction fit to an un-distorted mock with an increasing number of the largest eigenmodes (shown in the top-left corner of each plot) projected out of the resulting parameter covariance matrix. Points show the $k$-weighted deviation in the monopole $\Delta P(k,z)$ at $z = 2.4$ of the reduced data with errors calculated from the corresponding diagonal parameter covariance matrix elements. Curves show the fiducial model at $z = 2.4$ with (red) and without (blue, dashed) the projection distortion. Neither curve is a fit to the data points.}
\label{fig:pk-mock-fit-multi}
\end{center}
\end{figure}

\begin{figure}[htb]
\begin{center}
\includegraphics[width=6in]{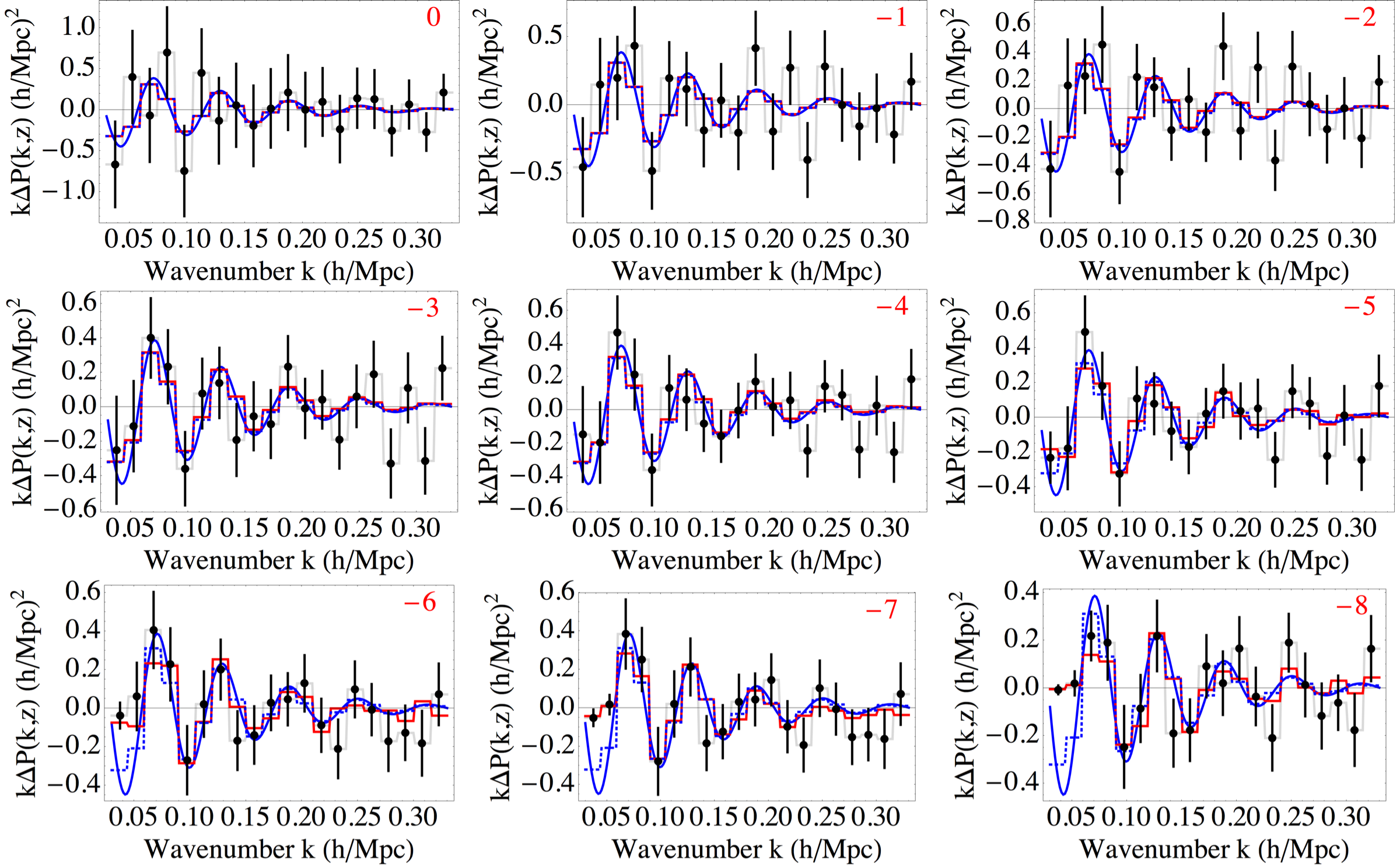}
\caption{Plots correspond to Figure~\ref{fig:pk-mock-fit-multi} but use a data-reduction fit that assumes identical $P(k)$ multipoles, resulting in 20 final parameters, as described in the text.}
\label{fig:pk-mock-fit}
\end{center}
\end{figure}

Figure~\ref{fig:interpolated-fit-cov} shows the ranked eigenvalues of the parameter covariance matrices for fits to data and the un-distorted mock dataset, and demonstrates that the data and mock have essentially the same parameter covariance structure. For the purposes of visualization, it is helpful to project out the largest eigenmodes of the parameter covariance matrix, which eliminates the largest sources of correlated errors but can also introduce a projection distortion of the reduced data (in addition to any distortion from the analysis method). The optimum number of modes to project out is a balance between minimizing correlated errors and minimizing projection distortion effects. Figs.~\ref{fig:xi-mock-fit}--\ref{fig:pk-mock-fit} show the results of projecting different numbers of eigenvalues on the monopole of our un-distorted mock, and allows us to identify an appropriate number of modes to filter for each data reduction fit: -14, -36, and -6, respectively.

\begin{figure}[htb]
\begin{center}
\includegraphics[width=6in]{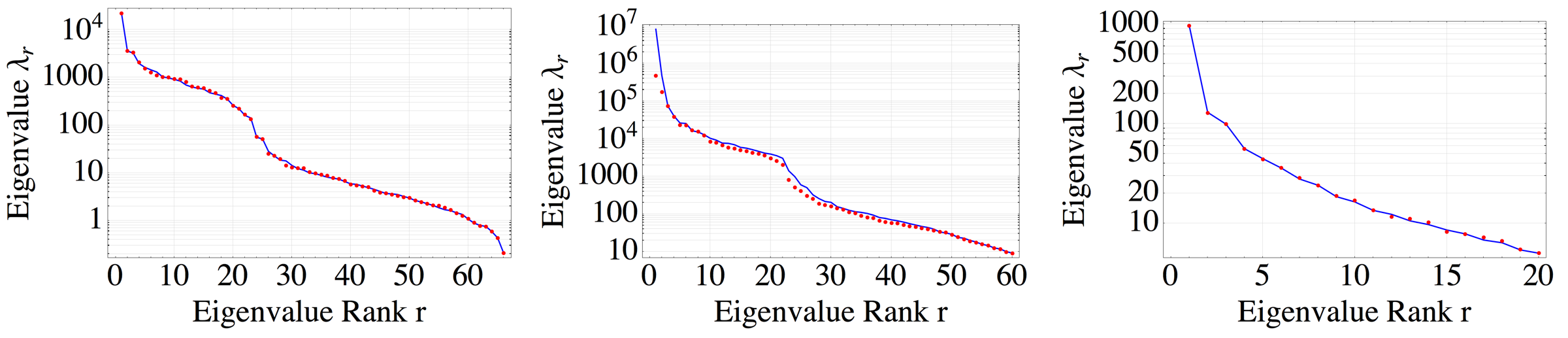}
\caption{Ranked eigenvalues of parameter covariance matrices for the three data reduction fits described in the text with, left to right, 66, 60, and 20 parameters, respectively. Red curves show eigenvalues for the fits to data and points show fits to a mock dataset without broadband distortion.}
\label{fig:interpolated-fit-cov}
\end{center}
\end{figure}

\begin{figure}[htb]
\begin{center}
\includegraphics[width=6in]{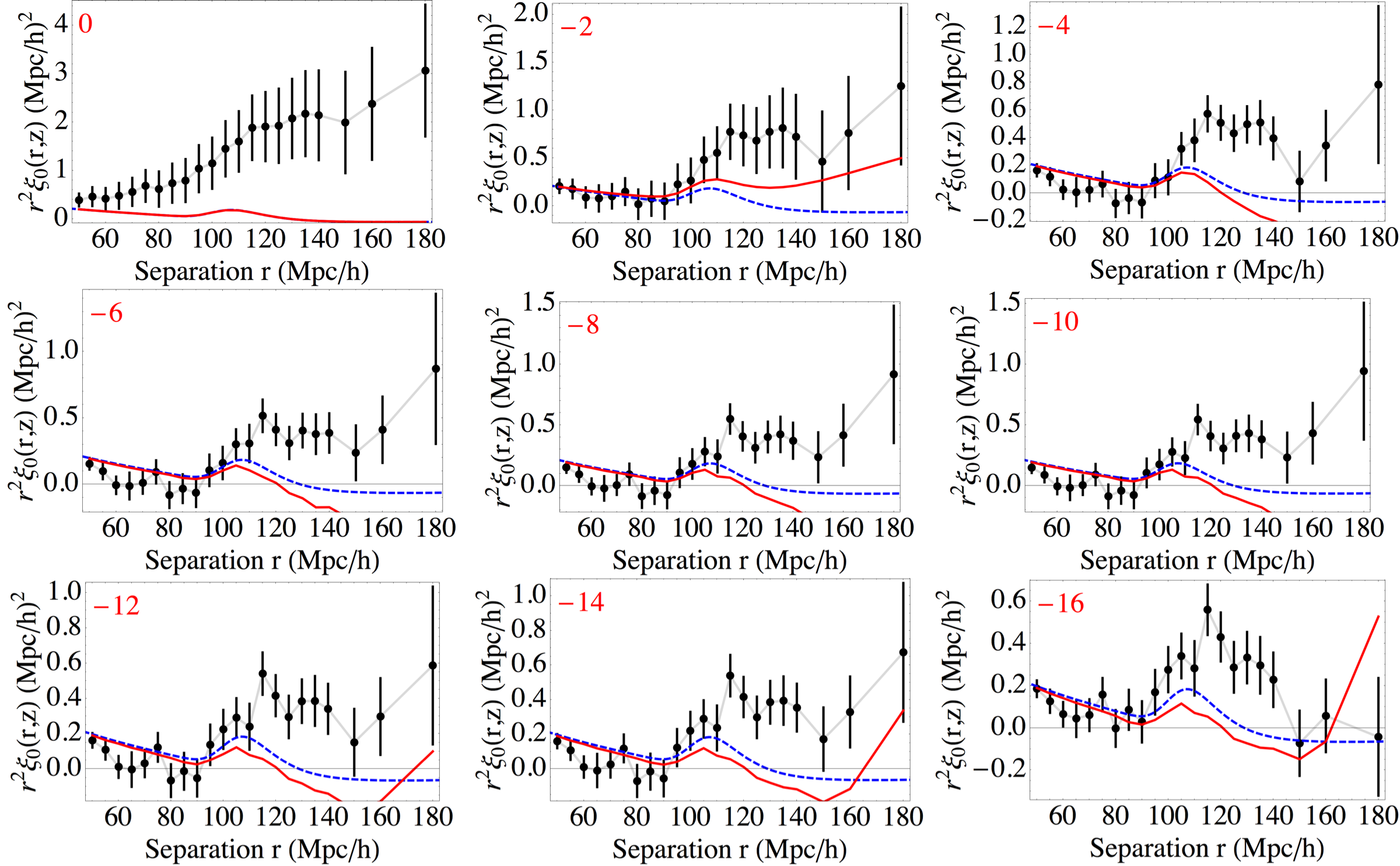}
\caption{Plots correspond to Figure~\ref{fig:xi-mock-fit} but are based on data instead of the un-distorted mock.}
\label{fig:xi-data-fit}
\end{center}
\end{figure}

\begin{figure}[htb]
\begin{center}
\includegraphics[width=6in]{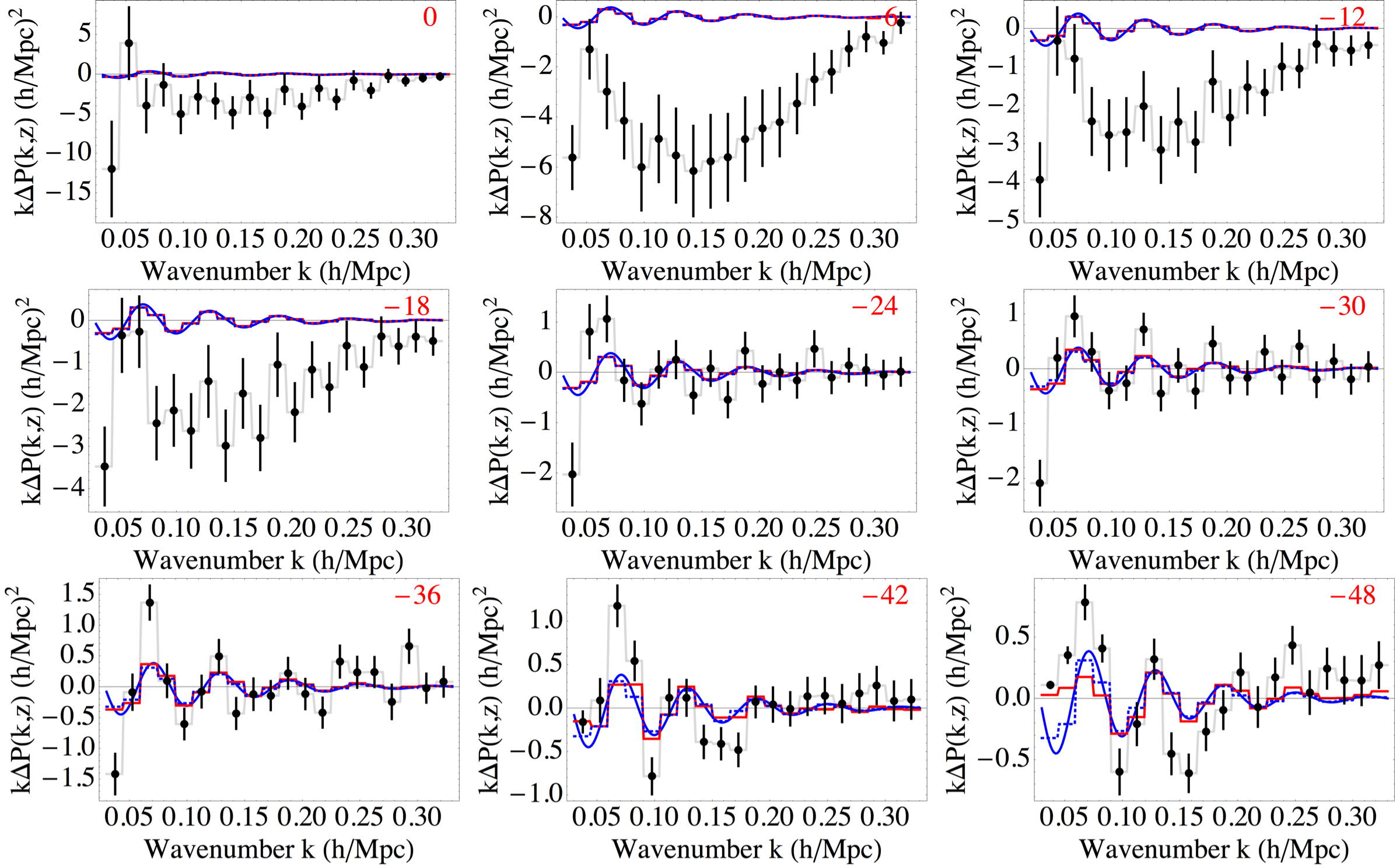}
\caption{Plots correspond to Figure~\ref{fig:pk-mock-fit-multi} but are based on data instead of the un-distorted mock.}
\label{fig:pk-data-fit-multi}
\end{center}
\end{figure}

\begin{figure}[htb]
\begin{center}
\includegraphics[width=6in]{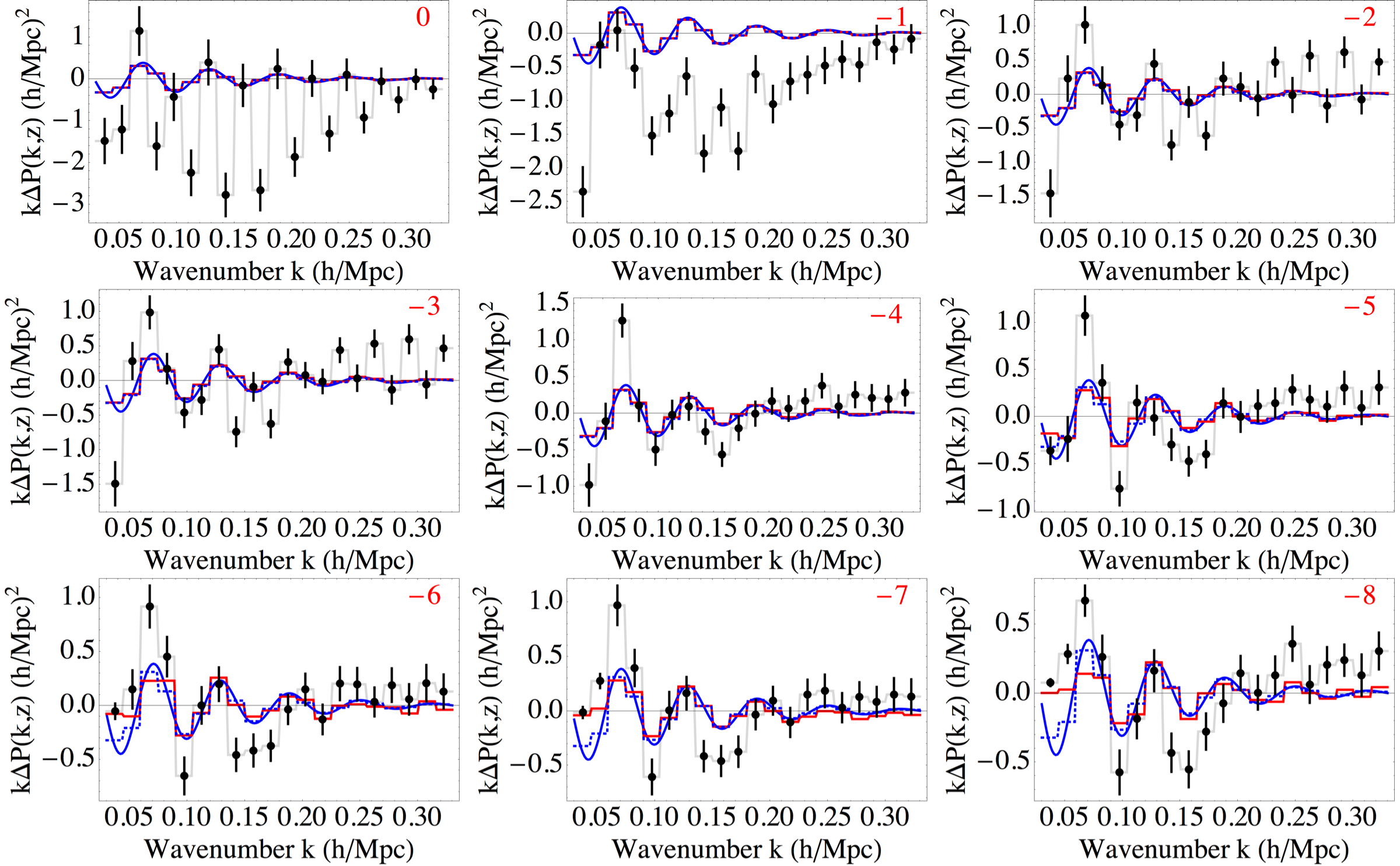}
\caption{Plots correspond to Figure~\ref{fig:pk-mock-fit} but are based on data instead of the un-distorted mock.}
\label{fig:pk-data-fit}
\end{center}
\end{figure}

Figs.~\ref{fig:xi-data-fit}--\ref{fig:pk-data-fit} show the same reductions as Figs.~\ref{fig:xi-mock-fit}--\ref{fig:pk-mock-fit}, but applied to data instead of un-distorted mocks. Comparing these, we conclude that the projection distortion effects are similar for each reduction. We also note that the analysis introduces significant broadband distortion that is most apparent (especially with the $r^2$ weighting) as a broad excess above 110 Mpc/h in the correlation-function monopole (Figure~\ref{fig:xi-mock-fit} with 14 modes projected out). Since this broadband distortion corresponds to low-$k$ modes, the $P(k)$ data reductions are relatively immune to it. Instead, they show an enhancement of the BAO oscillation signal relative to the fiducial model. Further study with additional data is needed to determine if this excess is a fortunate statistical fluctuation or if perhaps the signal is larger than expected.

\section{Discussion}
\label{sec:discussion}

In this paper we have described a near-optimal method for fitting correlations observed in the Lyman-$\alpha$ forest of high-redshift quasars to measure the properties of the baryon acoustic oscillation feature. We also explore the properties of the BOSS DR9 correlation estimates described in ref.~\cite{Slosar2013}, and provide the code and input files necessary to reproduce our main results.

Comparing with the Lyman-$\alpha$ fitting methods used in ref.~\cite{2012arXiv1211.2616B}, where correlations are modeled in the monopole and quadrupole at a single mean redshift, we have developed a fully three-dimensional fitting technique and included the effects of anisotropic non-linear broadening. These developments do not yield a significant improvement in the cosmological constraints that are possible with DR9, but provide more flexibility to study and quantify systematics and allow direct study the redshift evolution of parameters such as $b(z)$ and $\beta(z)$, and to accurately determine the effective redshift of our cosmological constraints.

The challenge of a three-dimensional analysis is that the resulting large covariance matrix is difficult to estimate and validate, especially with the limited number of mocks available for the DR9 Lyman-$\alpha$ forest. However, we have developed a novel method for internally testing and refining our large covariance matrix, as described in Section~\ref{sec:covariance-tests}. The resulting covariance matrix still has large correlations that visually obscure the BAO signal but have little impact on our ability to measure its properties. We describe data reductions that retain most of the cosmologically relevant information and show how the expected features are visually revealed when large eigenmodes are projected out.

In building models suitable for a three-dimensional analysis, we find that care is needed to isolate the BAO feature in the correlation function and describe an alternative to the ``no-wiggles'' approach of ref.~\cite{1998ApJ...496..605E} that is better suited for this purpose. When considering the line of sight ($\alpha_{\parallel}$) and transverse ($\alpha_{\perp}$) scale factors independently, we find that a first-order treatment is numerically inaccurate near the BAO peak, and identify the important second-order terms. We introduce anisotropic non-linear broadening using an approximation that is valid for a large range of $\beta$.
We also use a highly flexible parametrization of both multiplicative and additive broadband distortions to accomodate the expected systematics of the correlation-function estimates due to effects such as a biased continuum estimate. Finally, we demonstrate that information on the BAO feature in the Lyman-$\alpha$ forest arises primarily along the line of sight, as expected due to the large redshift-space distortion, but that the contributions from $H(z)$ and $D_A(z)$ can still be independently measured.

The first measurements of baryon acoustic oscillations in the Lyman-$\alpha$ forest~\cite{2012arXiv1211.2616B, Slosar2013} provide a new confirmation of our basic picture of cosmological evolution dominated by dark energy, but do not yet significantly constrain the parameters of the simplest cosmological models. With the SDSS-III high-redshift quasar sample expected to triple the DR9 statistics over the next 18 months, that situation is likely to change and will require near-optimal fitting methods and a flexible framework for careful study of systematics.

\section*{Acknowledgments}

Funding for SDSS-III has been provided by the Alfred P. Sloan
Foundation, the Participating Institutions, the National Science
Foundation, and the U.S. Department of Energy Office of Science. The
SDSS-III web site is \texttt{http://www.sdss3.org/}.

SDSS-III is managed by the Astrophysical Research Consortium for the
Participating Institutions of the SDSS-III Collaboration including the
University of Arizona, the Brazilian Participation Group, Brookhaven
National Laboratory, University of Cambridge, Carnegie Mellon
University, University of Florida, the French Participation Group, the
German Participation Group, Harvard University, the Instituto de
Astrofisica de Canarias, the Michigan State/Notre Dame/JINA
Participation Group, Johns Hopkins University, Lawrence Berkeley
National Laboratory, Max Planck Institute for Astrophysics, Max Planck
Institute for Extraterrestrial Physics, New Mexico State University,
New York University, Ohio State University, Pennsylvania State
University, University of Portsmouth, Princeton University, the
Spanish Participation Group, University of Tokyo, University of Utah,
Vanderbilt University, University of Virginia, University of
Washington, and Yale University. 

\def\apjl{ApJL} 
\def\aj{AJ} 
\def\apj{ApJ} 
\def\pasp{PASP} 
\def\spie{SPIE} %
\def\apjs{ApJS} 
\def\araa{ARAA} 
\def\aap{A\&A} 
\def\nat{Nature} 
\def\nar{New Astron. Rev.} 
\def\mnras{MNRAS} 
\def\jcap{JCAP} 
\def\prd{{Phys.~Rev.~D}}        

\bibliographystyle{JHEP}
\bibliography{cosmo,local}

\appendix
\section{Public Access to Data and Code}
\label{sec:public}

The software used to generate the results in this paper and ref.~\cite{Slosar2013} are publicly available at \texttt{http://github.com/baofit/}. We also provide instructions to install and run the software, together with the BOSS DR9 correlation-function estimates and configuration files necessary to reproduce our main results, at \texttt{http://darkmatter.ps.uci.edu/baofit/}. The software is written in C++ and uses MINUIT~\cite{minuit} for likelihood minimization. Data and configuration files are in plain text format.

\end{document}